\documentclass[10pt]{article}
\usepackage{bbm}
\usepackage[final]{graphics}
\usepackage{amsmath}
\usepackage{amsfonts,amsbsy}
\usepackage{amssymb}

\def\empile#1\over#2{\mathrel{\mathop{\kern 0pt#1}\limits_{#2}}}
\def\bs{\boldsymbol}

\newcommand{\slv}{\raise.15ex\hbox{$/$}\kern-.53em\hbox{$v$}}
\newcommand{\slF}{\raise.15ex\hbox{$/$}\kern-.53em\hbox{$F$}}
\newcommand{\slL}{\raise.15ex\hbox{$/$}\kern-.53em\hbox{$L$}}
\newcommand{\slP}{\raise.15ex\hbox{$/$}\kern-.53em\hbox{$P$}}
\newcommand{\slp}{\raise.15ex\hbox{$/$}\kern-.53em\hbox{$p$}}
\newcommand{\slq}{\raise.15ex\hbox{$/$}\kern-.53em\hbox{$q$}}
\newcommand{\slR}{\raise.15ex\hbox{$/$}\kern-.53em\hbox{$R$}}
\newcommand{\slQ}{\raise.15ex\hbox{$/$}\kern-.53em\hbox{$Q$}}
\newcommand{\slK}{\raise.15ex\hbox{$/$}\kern-.53em\hbox{$K$}}
\newcommand{\slk}{\raise.15ex\hbox{$/$}\kern-.53em\hbox{$k$}}
\newcommand{\slD}{\raise.15ex\hbox{$/$}\kern-.73em\hbox{$D$}}
\newcommand{\slC}{\raise.15ex\hbox{$/$}\kern-.53em\hbox{$C$}}
\newcommand{\slA}{\raise.15ex\hbox{$/$}\kern-.53em\hbox{$A$}}
\newcommand{\slSigma}{\raise.15ex\hbox{$/$}\kern-.53em\hbox{$\Sigma$}}
\newcommand{\slpartial}{\raise.15ex\hbox{$/$}\kern-.53em\hbox{$\partial$}}
\newcommand{\slcalP}{\raise.15ex\hbox{$/$}\kern-.63em\hbox{$\cal P$}}

\def\p{{\boldsymbol p}}
\def\q{{\boldsymbol q}}

\def\k{{\boldsymbol k}}

\def\x{{\boldsymbol x}}
\def\y{{\boldsymbol y}}

\def\z{{\boldsymbol z}}

\catcode`\@=11

\newcount\@tempcntc
\def\@citex[#1]#2{\if@filesw\immediate\write\@auxout{\string\citation{#2}}\fi
  \@tempcnta\z@\@tempcntb\m@ne\def\@citea{}\@cite{%
        \@for\@citeb:=#2\do%
    {\@ifundefined{b@\@citeb}%
        {\@citeo\@tempcntb\m@ne\@citea%
                \def\@citea{,\penalty\@m\ }{\bf ?}\@warning%
                {Citation `\@citeb' on page \thepage \space undefined}}%
        {\setbox\z@\hbox{\global\@tempcntc0\csname b@\@citeb\endcsname\relax}
     \ifnum\@tempcntc=\z@ \@citeo\@tempcntb\m@ne%
       \@citea\def\@citea{,\penalty\@m}%
       \hbox{\csname b@\@citeb\endcsname}%
     \else%
      \advance\@tempcntb\@ne%
      \ifnum\@tempcntb=\@tempcntc%
      \else\advance\@tempcntb\m@ne\@citeo%
      \@tempcnta\@tempcntc\@tempcntb\@tempcntc\fi\fi}}\@citeo}{#1}}%

\def\@citeo{\ifnum\@tempcnta>\@tempcntb\else\@citea
  \def\@citea{,\penalty\@m}%
  \ifnum\@tempcnta=\@tempcntb\the\@tempcnta\else
   {\advance\@tempcnta\@ne\ifnum\@tempcnta=\@tempcntb \else
\def\@citea{--}\fi
    \advance\@tempcnta\m@ne\the\@tempcnta\@citea\the\@tempcntb}\fi\fi}

\catcode`\@=12

\begin{document}

\title{\bf Formulation of the Schwinger mechanism
 in classical statistical field theory}
\author{Fran\c cois Gelis${}^{(a)}$, Naoto Tanji${}^{(a,b)}$}

\maketitle
 \begin{center}
   \begin{itemize}
\item[{a.}] Institut de Physique Th\'eorique (URA 2306 du CNRS)\\
   CEA/DSM/Saclay, 91191 Gif-sur-Yvette Cedex, France
\item[{b.}] High Energy Accelerator Research Organization (KEK)\\
 1-1 Oho, Tsukuba, Ibaraki 305-0801, Japan
\end{itemize}
 \end{center}

\begin{abstract}
  In this paper, we show how classical statistical field theory
  techniques can be used to efficiently perform the numerical
  evaluation of the non-perturbative Schwin\-ger mechanism of particle
  production by quantum tunneling. In some approximation, we also
  consider the back-reaction of the produced particles on the external
  field, as well as the self-interactions of the produced particles.
\end{abstract}

\section{Introduction}
The Schwinger mechanism \cite{Schwi2} (see \cite{Dunne1} for a
comprehensive review) is a phenomenon by which charged particles,
e.g. electron-positron pairs, are produced spontaneously from an
external electrical field. This phenomenon is non-perturbative since
pairs can be produced even from a static electrical field, something
which is forbidden at any finite order of perturbation theory by
simple kinematical arguments. It is also a purely quantum phenomenon,
whose probability goes to zero in the classical limit
$\hbar\to0$. Loosely speaking, $e^+e^-$ vacuum fluctuations are
promoted to on-shell real particles by picking energy to the
electrical field -- which can be viewed as a kind of quantum tunneling
process.

In Quantum Electrodynamics, the probability for particle production by
the Schwinger mechanism is of the order of $\exp(-\pi m^2 / eE)$, for
particles of mass $m$ and electrical charge $e$, in a field $E$. For
fields one may realistically create in experiments, and taking even
the lightest charged particle, the electron, this probability is so
small (mostly due to the fact that the coupling constant $e$ is small)
that this phenomenon has remained elusive in all laboratory
experiments so far (the typical electrical field necessary to make the
production of an electron-positron pair likely is of the order of
$E\sim m^2/e\sim 10^{18}$~V/m).

The subject of pair production by the Schwinger mechanism is also
relevant in the context of Quantum Chromodynamics and strong
interactions, since the strong coupling constant $g$ is much
larger. It is for instance an important ingredient in hadronization
models such as the Lund string model \cite{AnderGIS1}, where the
breaking of a ``string'' made of a color electrical field into
quark-antiquark pairs leads to meson production. It is also an
ingredient in several phenomenological models of heavy ion collisions,
e.g.  \cite{CasheNN1,GlendM1,BiroNK1,KajanM1,GatofKM1,WangW1}.

It may also be a relevant mechanism of particle production in the
Color Glass Condensate framework (see
\cite{IancuLM3,IancuV1,Lappi6,GelisIJV1,Gelis15}), that is commonly
used in the description of the first stages of hadronic or nuclear
collisions at high energy. In this effective theory, the fast partons
--mostly gluons at high energy-- of the two colliding projectiles act
as a static classical color source. The gluon occupation number, and
therefore also this color source, increases with energy. Eventually,
when the gluon occupation becomes of order of the inverse strong
coupling $1/g^2$, nonlinear effects that tame this growth become
important -- an effect known as gluon saturation
\cite{GriboLR1,MuellQ1,BlaizM2}. In this regime, the color source
corresponding to the fast partons is of order $1/g$, and therefore it
creates fields that are themselves of order $1/g$.  The probability of
pair creation by such a strong field is not suppressed since $gE$ can
be large, unlike in QED. In \cite{FukusGL1}, it has been argued that
the CGC framework at next-to-leading order (1-loop) includes the
contribution of the Schwinger mechanism to particle production.

The Color Glass Condensate provides a semi-classical description of
the underlying dynamics: at leading order (tree level), observables
are computed by solving classical field equations of motion. This
power counting is justified by the large occupation numbers and large
fields that characterize the saturation regime, that allow one to
neglect the non-commuting nature of the quantum fields and treat them
as classical.  The color fields obtained at leading order --the
Glasma~\cite{LappiM1,Gelis15}-- are non-perturbatively large, of order
$1/g$, and can thus lead to a large pair production.  At
next-to-leading order (1-loop), one can formulate observables in terms
of small perturbations of this classical
field~\cite{GelisV2,GelisLV3}, that obey linearized (and still
classical) equations of motion. Equivalently, these 1-loop corrections
can be calculated, and resummed, by computing a classical path
integral where one sums over a Gaussian ensemble of initial conditions
for the classical field encountered at leading order. This approach,
sometimes called classical statistical field theory, has been employed
in a number of problems in cosmology \cite{PolarS1,Son1,MichaT1}, cold
atom physics \cite{Norri1,NorriBG1}, and more recently in computations
related to the thermalization of the quark-gluon plasma in heavy ion
collisions
\cite{DusliEGV1,EpelbG1,DusliEGV2,BergeSSS1,BergeS6,BergeSS3}.

The traditional method of computing the Schwinger mechanism has been
to obtain it from the imaginary part of the Heisenberg-Euler
Lagrangian \cite{HeiseE1}, which in more modern quantum field theory
language corresponds to calculating the imaginary part of the 1-loop
effective action. It has also been derived in the framework of kinetic
theory~\cite{SchmiBRSP1,SchmiBRPS1,BlascDRS1}. In
Refs.~\cite{Tanji1,Tanji2,Tanji3}, the Schwinger mechanism was
computed from the Bogoliubov transformation that maps the
creation-annihilation operators at $t=+\infty$ onto those at
$t=-\infty$. This requires that one solves the linearized equations of
motion in order to obtain the time evolution of the mode functions
(i.e. modes that start as plane waves in the remote past, and are
distorted by their propagation over the external field).

Motivated by the applications of the classical statistical method to
the CGC framework, we show in the present paper how the Schwinger
mechanism can be calculated in classical statistical field theory,
despite being an intrinsically quantum phenomenon.  In order to keep
the formalism as light as possible, we consider scalar electrodynamics
instead of QCD.  Note that a similar approach has been used in the
case of fermion production in ref.~\cite{HebenBG1,SaffiT1,SaffiT2},
following an idea of ref.~\cite{BorsaH1} to simulate fermions
efficiently on a lattice.

In the section \ref{sec:spectrum-LO}, we describe the model and
discuss two methods of calculating the spectrum of produced particles
at leading order; first in a rather standard quantum field theory
formulation, and secondly as a classical path integral.  In the
section \ref{sec:num-1loop}, we describe the lattice formulation of
this calculation in order to evaluate the particle spectrum
numerically, and then we compare the results of the classical
statistical approach to the results obtained by the direct
calculation of the 1-loop diagram, in order to show that they are
indeed equivalent. We improve this calculation in the section
\ref{sec:br} in order to include the back-reaction of the produced
particle pairs on the electrical field. Indeed, this screening effect
is crucial for proper energy conservation. The self-interactions among
the produced particles, that are crucial for the eventual
thermalization of the system, are considered in the section
\ref{sec:self}. Since the issue of thermalization in classical
statistical field theory has been addressed elsewhere, we focus here
on the issue of mass renormalization, which plays a crucial role in
the Schwinger mechanism due to its extreme sensitivity on the mass of
the particles being produced.  Finally, the section \ref{sec:conc}
contains some concluding remarks, while details about the mass
renormalization are relegated to the appendix \ref{app:tadpole}.

\section{Single inclusive spectrum at leading order}
\label{sec:spectrum-LO}
\subsection{Scalar QED model}
Let us consider the case of a complex scalar field $\phi$ with U(1)
symmetry, minimally coupled to an Abelian vector field $A^\mu$. The
vector field may be coupled to an external source $J^\mu$ that drives
it to a non-perturbatively large value.
The classical Lagrangian of this model is
\begin{eqnarray}
&&
{\cal L}\equiv -\frac{1}{4}F_{\mu\nu}F^{\mu\nu}+
(D_\mu \phi)(D^\mu\phi)^* -m^2\phi^*\phi-V(\phi\phi^*)+J^\mu_{\rm ext} A_\mu\nonumber\\
&&
F^{\mu\nu}= \partial^\mu A^\nu-\partial^\nu A^\mu\quad,\quad
D^\mu\equiv\partial^\mu -ie A^\mu\; ,
\end{eqnarray}
where $e$ is the electrical charge of the scalar particles described
by the field $\phi$.  We have not specified for now the
self-interaction potential $V$ of the scalar field, except for the
fact that it depends only on the U(1) invariant $\phi\phi^*$. A
typical example of such a potential would be a quartic interaction,
\begin{equation}
V(\phi\phi^*) = \frac{\lambda}{4}(\phi\phi^*)^2\; ,
\end{equation}
where $\lambda$ sets the strength of the self-interactions.

The external source $J^\mu_{\rm ext}$ can produce a non-trivial gauge potential,
which in turn may produce scalar particles. Assuming that the initial
state of the system is the vacuum, the inclusive
spectrum\footnote{This observable should not be confused with the
  probability $P_1$ of producing exactly one particle-antiparticle
  pair, that would be obtained from the matrix element $\big<0_{\rm
    out}\big|\phi(x)\big|0_{\rm in}\big>$. The average number of
  produced particles (i.e. the integral over $\p$ of $dN_1/d^3\p$) is
  related to the probabilities $P_n$ by $N_1=\sum_n nP_n$. $N_1$ is
  usually easier to calculate than the individual $P_n$, thanks to
  simplifications related to the completeness of the set of possible
  final states.} of scalar particles is given by the following formula
in terms of the 2-point correlation function of the field $\phi$,
\begin{equation}
  \frac{dN_1}{d^3\p}
  =
  \frac{1}{(2\pi)^3 2E_\p}
  \int d^4x d^4y\;
  e^{-ip\cdot(x-y)}
  \;
  (\square_x+m^2)(\square_y+m^2)
  \,
  \big<0_{\rm in}\big|\phi^{\dagger} (x)\phi(y)\big|0_{\rm in}\big>\; ,
\label{eq:LSZ1}
\end{equation}
where $E_\p^2\equiv \p^2+m^2$.  This expression is simply the
Lehmann-Symanzik-Zimmermann reduction formula for the expectation
value of the number operator $a^\dagger(\p)a(\p)$. Although this form
of the reduction formula involves 4-dimensional integrals over the
entire space-time, it can also be written in terms of purely spatial
integrals thanks to the identity
\begin{equation}
  \int\! d^4x\, e^{ip\cdot x}\;(\square_x+m^2) \phi(x)
  =
  \!\int\! d^3\x\, e^{ip\cdot x}\;\big[\dot\phi(x^0,\x)
-iE_\p\phi(x^0,\x)\big]^{x^0=+\infty}_{x^0=-\infty}\; ,
\label{eq:4->3}
\end{equation}
where $[A(x^0)]^b_a\equiv A(b)-A(a)$ and the dot denotes a time
derivative. Moreover, the lower boundary $x^0\to-\infty$ does not
contribute since we assume that the initial state is
empty\footnote{The contribution of the lower boundary is in fact
  $\big<0_{\rm in}\big|a^\dagger_{\rm in}(\p)a_{\rm in}(\p)\big|0_{\rm
    in}\big>=0$.}. Thus, eq.~(\ref{eq:LSZ1}) can also be written
as\footnote{At this stage, the object $\phi$ is still an operator, and
  one should keep in mind that the commutator $[\dot\phi,\phi]$ is not
  zero.}
\begin{eqnarray}
  \frac{dN_1}{d^3\p}
  &=&\lim_{t\to+\infty}
  \frac{1}{(2\pi)^3 2E_\p}
  \int d^3\x d^3\y\;
  e^{+i\p\cdot(\x-\y)}
  \nonumber\\
  &&\quad\quad\times
  \big<0_{\rm in}\big|(\dot\phi^{\dagger} (t,\x)
+iE_\p\phi^{\dagger} (t,\x))(\dot\phi(t,\y)
-iE_\p\phi(t,\y))\big|0_{\rm in}\big>\; .
\label{eq:LSZ2}
\end{eqnarray}
On may remove the limit $t\to\infty$ in this formula, and interpret
its result as the particle spectrum at the time $t$. However, one has
to keep in mind that this interpretation cannot be completely
rigorous: strictly speaking, the particles need to be free and
on-shell in order for their number to be a well defined concept, which
takes an infinite time.

In addition, this formula for the spectrum assumes that the gauge
potential vanishes when $t\to+\infty$. However, even if the electrical
and magnetic fields are made to vanish in this limit, the gauge
potential itself could be a non-zero pure gauge
$A^\mu\equiv\partial^\mu \chi$. In this case, we need to perform in
eq.~(\ref{eq:LSZ2}) the replacement
\begin{equation}
e^{-i\p\cdot\x}\quad\to\quad e^{-i\p\cdot\x}\,e^{ie\chi(x)}\; .
\end{equation}
(In words, we must gauge transform the free plane waves that are used
in the Fourier decomposition of the fields). Note that this
replacement is also required in order to have a gauge invariant
spectrum. In the particular case where the gauge potential ${\bs A}$
is spatially homogeneous, this substitution can also be written as
\begin{equation}
e^{-i\p\cdot\x}\quad\to\quad e^{-i(\p+e{\bs A})\cdot\x}\; ,
\end{equation}
and we recognize now the well known difference between the kinetic and
canonical momenta\footnote{%
One can find a similar argument on the LSZ formula under 
a background gauge field in Refs.~\cite{Dinu1,Kibble1}. 
}
\begin{equation}
\p_{\rm cano}=\p_{\rm kin}+e{\bs A}\; .
\end{equation}
By extension, we will also perform this substitution in order to
define the particle spectrum in the presence of a homogeneous
electrical field (i.e. when the gauge potential is not a pure
gauge). Again, one must keep in mind that the concept of particle
number in the presence of a non-trivial background field is not
rigorously defined, since the particles we are trying to count are not
free particles.

\subsection{Spectrum at Leading Order}
\begin{figure}[htbp]
\begin{center}
\resizebox*{6cm}{!}{\includegraphics{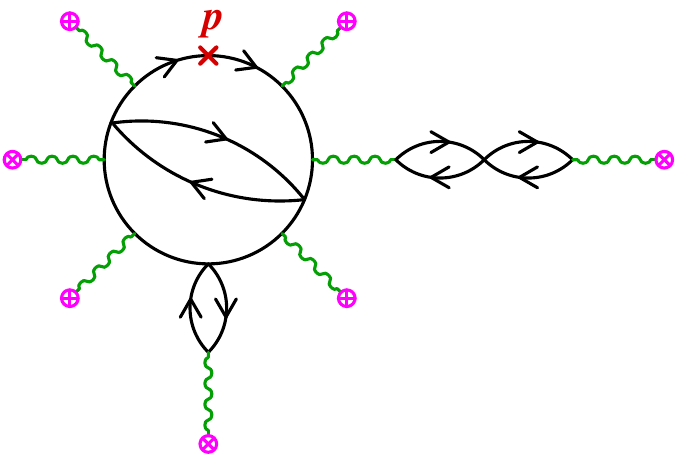}}
\end{center}
\caption{\label{fig:gen}Generic contribution to the inclusive spectrum
  of produced scalar particles. The wavy lines represent the photons,
  while the solid black lines are scalars. The crosses terminating the
  photon lines represent the source $J^\mu_{\rm ext}$.}
\end{figure}

A typical graph contributing to the spectrum is shown in the figure
\ref{fig:gen}.
The spectrum can be organized as a triple series expansion, in powers
of the electromagnetic coupling $e$, of the self-coupling $\lambda$,
and of the external source $J^\mu_{\rm ext}$.

\begin{figure}[htbp]
\begin{center}
\resizebox*{3.5cm}{!}{\includegraphics{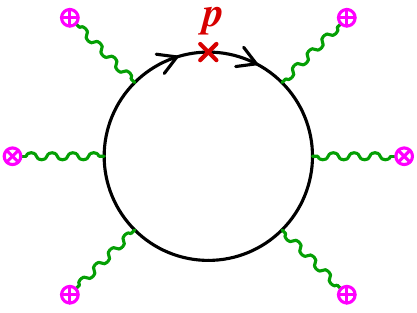}}
\end{center}
\caption{One of
  graphs contributing to the inclusive spectrum
  at leading order.
  The leading order consists of all such graphs which have
  arbitrary number of external lines connected to the external 
  source\protect\footnotemark.
  }
\label{fig:LO}
\end{figure}

When the external source $J^\mu_{\rm ext}$ is large, possibly of order
$J_{\rm ext}\sim {\cal O}(e^{-1})$, one may expect non-perturbative
effects such as the Schwinger mechanism to become important. 
In order to compute these contributions which are usually non analytic 
in $e$, it is necessary to treat exactly the external source 
$J^\mu_{\rm ext}$.
In the rest of the paper, we call {\sl Leading Order}
the result of this treatment, 
in which we include all powers of $e$
accompanied by a power of $J^\mu_{\rm ext}$, but no further
corrections in $e^2$ or in $\lambda$.  Therefore, the graphs that
contribute to the spectrum at leading order are considerably simpler
(see the figure \ref{fig:LO}), since they have only one scalar loop
dressed by insertions of a photon directly connected to the external
source $J^\mu_{\rm ext}$.
\footnotetext{%
If one computes such diagrams individually and then sum them up, 
the correct nonperturvative contribution cannot be obtained. 
(The integration with respect to the loop momentum and the summation over
the number of the external lines do not commute.) 
To get the correct answer, we have to solve the field equation for $\phi$ 
treating the interaction with the background field exactly.
This situation is similar to considering the Taylor expansion of the function
$f(z)=e^{-1/z}$ around its non-analytic point $z=0$. 
}

At this level of approximation, one can treat the gauge field
attached to the scalar loop is a classical field ${\cal A}^\mu$ that
obeys the classical Maxwell's equations,
\begin{equation}
\partial_\mu{\cal F}^{\mu\nu}=J^\nu_{\rm ext}\; ,
\end{equation}
and the equation for $\phi$ is linear with respect to quantum fields:
\begin{equation}
({\cal D}_\mu{\cal D}^\mu+m^2)\,\phi=0\; ,
\end{equation}
where ${\cal D}_\mu\equiv\partial_\mu-ie{\cal A}_\mu$ is the covariant
derivative constructed with the classical background field.
A solution of this equation can be expanded by normal modes as
\begin{equation} \label{eq:modeexp}
\phi (x) = \int\frac{d^3\q}{(2\pi)^3 2E_\q}
\left[ \varphi_\q (x) a_\text{in} (\q) 
+\varphi_\q^* (x) b_\text{in} ^\dagger (\q) \right] , 
\end{equation}
where $a_\text{in} (\q) $ and $a_\text{in} (\q) $ are the annihilation 
operator for particle and antiparticle, respectively, 
and follow the commutation relations
\begin{eqnarray}
  \left[ a_\text{in} (\q), a_\text{in}^\dagger (\q') \right] 
  = \left[ b_\text{in} (\q), b_\text{in}^\dagger (\q') \right]
  =(2\pi)^3 2E_\q \delta(\q-\q') \; . \label{eq:cr}
\end{eqnarray}
The positive frequency mode function $\varphi_{\q} (x)$ follows
\begin{equation}
({\cal D}_\mu{\cal D}^\mu+m^2)\varphi_{\q}(x)=0\quad,\quad
\lim_{x^0\to-\infty}\varphi_{\q}(x)=e^{-iq\cdot x}\; .
\end{equation}
The 2-point correlation function that enters in the spectrum is
expressible as follows
\begin{eqnarray}
\big<0_{\rm in}\big|\phi^\dagger (x)\phi(y)\big|0_{\rm in}\big>_{_{\rm LO}}
=
\int\frac{d^3\q}{(2\pi)^3 2E_\q}\;
\varphi_{\q}(x)\varphi_{\q}^* (y) \label{eq:LO1} 
\end{eqnarray}

At this point, the calculation of the scalar particle yield at leading
order has been recasted into a purely classical calculation, where one
needs to solve the classical equation of motion $({\cal D}_\mu{\cal
  D}^\mu+m^2)\varphi=0$ for each of the scalar mode functions
$\varphi_{\q}$. In the special case of a static electrical field,
eqs.~(\ref{eq:LO1}) are equivalent to the classic result of Schwinger.
Note that eqs.~(\ref{eq:LO1}) do not imply that the particle spectrum
at leading order is a classical quantity. Indeed, it is well known in
the case of a static electrical field that the particles are produced
by a quantum tunneling phenomenon, whose probability goes to zero if
$\hbar\to 0$. Instead, eqs.~(\ref{eq:LO1}) should be viewed as an
example of the general property that one-loop quantities can be
written as quadratic forms in terms of fields that obey linearized
classical equations of motion.

\begin{figure}[htbp]
\begin{center}
\resizebox*{3.5cm}{!}{\includegraphics{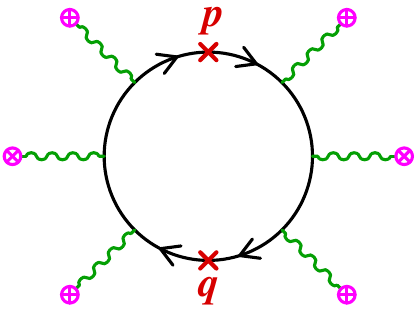}}
\end{center}
\caption{\label{fig:LO1}Diagrammatic representation of eqs.~(\ref{eq:LO1}).}
\end{figure}

A crucial aspect of this formulation is that these mode functions are
specified by retarded boundary conditions in the remote past, where
they behave as free plane waves. Using the identity (\ref{eq:4->3})
and the fact that in the limit $x^0\to-\infty$, we have
$\varphi_{\q}^*(x)=e^{iq\cdot x}$, we simply have
\begin{equation}
  \int\! d^4x\, e^{ip\cdot x}\,(\square_x+m^2) \varphi_{\q}^*(x)
  =
  \lim_{x^0\to+\infty}
  \!\int\! d^3\x\, e^{ip\cdot x}\;\big[\dot\varphi_{\q}^* (x^0,\x)
-iE_\p\varphi_{\q}^* (x^0,\x)\big]\, .
\end{equation}
Thus, as illustrated in the figure \ref{fig:LO1}, the physical
interpretation of eqs.~(\ref{eq:LO1}) is that in order to obtain the
spectrum of produced particles, one should start in the remote past
with negative energy plane waves (that are equivalent, by crossing
symmetry, to having a positive energy antiparticle in the final
state), that subsequently evolve over the classical gauge field ${\cal
  A}^\mu$, and are projected at the final time on a positive energy
plane wave. The momentum $\q$ of the incoming plane wave can be
interpreted as the momentum of the antiparticle that must be produced
along with the observed particle of momentum $\p$, and therefore
should be integrated out in order to obtain the particle spectrum.

This formulation provides an explicit numerical method for computing
the yield at leading order for a general background field (for which
it is not possible to solve analytically the equation of motion for
$\varphi_{\q}$): discretize space and solve numerically the
classical equation of motion for each momentum $\q$ of the reciprocal
lattice. This is however a computationally expensive method, since
this computation scales like the square of the number of lattice
points. More precisely, the computation time scales as
\begin{equation}
N_t \times N_{\rm latt}^2\; ,
\label{eq:t1}
\end{equation}
where $N_{\rm latt}$ is the number of lattice points and $N_t$ the
number of time-steps used in solving the equations of motion.

\subsection{Reformulation as a Gaussian functional integral}
It is however possible to formulate the particle spectrum in an
alternative way, that allows a more efficient computation based on a
Monte-Carlo sampling in a functional space that we shall specify
shortly.  Since the evolution of the mode functions is causal, knowing
them, as well as their first order time derivative\footnote{This is
  necessary because the equation of motion for $\varphi$ contains
  second order time derivatives.}, on some Cauchy surface $\Sigma$
(for instance, any surface of constant time $x^0$) is sufficient to
fully determine their subsequent evolution. To illustrate this, let us
consider the following functional
\begin{eqnarray}
&&
G_{xy}[\varphi_0,\pi_0]\equiv\varphi^*(x)\varphi(y)\nonumber\\
&&
({\cal D}_\mu{\cal D}^\mu+m^2)\,\varphi=0\quad,\quad
\varphi(t_0,\x)=\varphi_0(\x)\;,\;
\dot\varphi(t_0,\x)=\pi_0(\x)\; .
\end{eqnarray}
$G_{xy}[\varphi_0,\pi_0]$ is the product at the points $x$ and $y$ of
the classical solution $\varphi$ whose initial conditions at the time
$t_0$ are given by the functions $\varphi_0(\x),\pi_0(\x)$. 
Let us introduce a Gaussian average 
over initial values $\varphi_0(\x),\pi_0(\x)$ as
\begin{equation}
\big< \mathcal{O} \left[\varphi_0 ,\pi_0 \right] \big>
\equiv \int [D\varphi_0 D\pi_0]\;W[\varphi_0,\pi_0]\;\mathcal{O}[\varphi_0,\pi_0] ,
\end{equation}
with $W[\varphi_0,\pi_0]$ being the Gaussian kernel which is characterized by the
following 2-point correlation functions~:
\begin{align}
\big<\varphi_0^*(\x)\varphi_0(\y)\big> &= 
  \frac{1}{2}\int \frac{d^3\q}{(2\pi)^3 2E_\q}\;
  \Big[ \varphi_{\q}^*(t_0,\x)\varphi_{\q}(t_0,\y) 
  +\varphi_{\q}(t_0,\x)\varphi_{\q}^*(t_0,\y) \Big] 
  \nonumber\\
  \big<\pi_0^*(\x)\pi_0(\y)\big> &=
  \frac{1}{2}\int \frac{d^3\q}{(2\pi)^3 2E_\q}\;
  \Big[ \dot\varphi_{\q}^*(t_0,\x)\dot\varphi_{\q}(t_0,\y) 
  +\dot\varphi_{\q}(t_0,\x)\dot\varphi_{\q}^*(t_0,\y) \Big] 
  \label{eq:init-corr} \\
  \big<\varphi_0(\x)\pi_0(\y)\big> &=
  0\; . \nonumber
\end{align}
From these definitions, the following equation is trivially obtained:
\begin{equation}
\big< G_{xy}[\varphi_0,\pi_0] \big> 
=\frac{1}{2}\int \frac{d^3\q}{(2\pi)^3 2E_\q}\;
\Big[ \varphi_{\q}(x)\varphi_{\q}^*(y) 
 +\varphi_{\q}(y)\varphi_{\q}^*(x) \Big] \; .
\end{equation}
Further using Eq.~\eqref{eq:LO1}, we get
\begin{equation}
\big< G_{xy}[\varphi_0,\pi_0] \big> 
=\frac{1}{2}\big<0_{\rm in}\big|\phi^\dagger (x)\phi(y)+\phi(y)\phi^\dagger(x)\big|0_{\rm in}\big>_{_{\rm LO}}\; .
\label{eq:LSZ3}
\end{equation}
In other words, this procedure gives almost the expectation value we
need in order to compute the particle spectrum, except for the
ordering between the two field operators. In the form (\ref{eq:LSZ2})
of the reduction formula, the two field operators are evaluated at
equal times. Therefore, $\phi^\dagger(x)$ and $\phi(y)$ commute, but not
$\dot\phi^\dagger(x)$ and $\phi(y)$.  In fact, if we used the expectation
value (\ref{eq:LSZ3}) in the LSZ formula, we would get the expectation
value of the operator $(a^\dagger(\p)a(\p)+a(\p)a^\dagger(\p))/2$
instead of $a^\dagger(\p)a(\p)$. This discrepancy is easy to fix by
using the commutation relation \eqref{eq:cr}.
One can obtain the leading order spectrum by calculating the expectation
value (\ref{eq:LSZ3}), and then by subtracting $V/2$ particles where
$V$ is the volume\footnote{The volume arises from a $\delta({\bs 0})$
  in momentum space.} of the system. More precisely,
\begin{equation}
\left.\frac{dN_1}{d^3\p}\right|_{_{\rm LO}}
  =
-\frac{V}{2}
+\int [D\varphi_0 D\pi_0]\;W[\varphi_0,\pi_0]\; F_\p[\varphi_0,\pi_0]\; ,
\label{eq:LO2}
\end{equation}
where 
\begin{eqnarray}
&&
F_\p[\varphi_0,\pi_0]
\equiv
\frac{1}{(2\pi)^3 2E_\p} \left|\Phi_\p[\varphi_0,\pi_0]\right|^2
\nonumber\\
&&
\Phi_\p[\varphi_0,\pi_0]
=
\lim_{x^0\to+\infty}
  \int d^3\x\; e^{ip\cdot x}\;\big[\dot\varphi(x^0,\x)
-iE_\p\varphi(x^0,\x)\big]\nonumber\\
&&({\cal D}_\mu{\cal D}^\mu+m^2)\,\varphi=0\quad,\quad
\varphi(t_0,\x)=\varphi_0(\x)\;,\;
\dot\varphi(t_0,\x)=\pi_0(\x)\; .
\end{eqnarray}
It is easy to check that if we carry out this procedure in the vacuum
(i.e. with ${\cal D}_\mu=\partial_\mu$ in the equation of motion of
the scalar field), we obtain $dN_1/d^3\p=0$. The subtraction of the
term $V/2$ is crucial for that.

This reformulation of the spectrum at leading order can be illustrated
diagrammatically as follows\footnote{Although in this illustration we
  have put all the interactions with the background field in the
  functional $F_\p[\varphi_0,\pi_0]$, this is not mandatory. If the
  time $t_0$ is chosen larger than the time at which the background
  field is switched on, then the Gaussian distribution
  $W[\varphi_0,\pi_0]$ also depends on the background field.},
\setbox1\hbox to
30mm{\resizebox*{30mm}{!}{\includegraphics{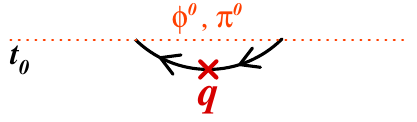}}} \setbox2\hbox
to 30mm{\resizebox*{30mm}{!}{\includegraphics{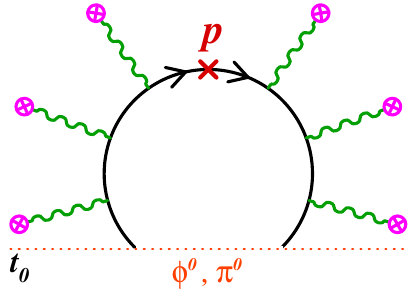}}}
\setbox3\hbox to 30mm{\resizebox*{30mm}{!}{\includegraphics{sp_LO1}}}
\begin{equation}
\raise -10mm\box3
\;=\;
\underbrace{\raise -5mm\box1}_{W[\varphi_0,\pi_0]}
\;\otimes\;
\underbrace{\raise -5mm\box2}_{F_\p[\varphi_0,\pi_0]}\; .
\label{eq:diam}
\end{equation}
The new representation amounts to slicing the scalar loop at a certain
time $t_0$, and to assign to the functional $F_p[\varphi_0,\pi_0]$ the
time evolution for $x^0>t_0$, and to the Gaussian distribution
$W[\varphi_0,\pi_0]$ the part of the evolution at $x^0<t_0$. The
Gaussian average over the fields $\varphi_0,\pi_0$ is the ``glue''
that reconstructs the loop, since it amounts to connecting pairwise
the two open endpoints of $F_p[\varphi_0,\pi_0]$. Note that the choice
of the time $t_0$ is not important, since the left hand side does not
depend on this choice. In fact, instead of slicing the loop at a fixed
time $t_0$, the separation in two factors could have been made on any
locally space-like surface. In practical implementations, it is best
to perform the separation at a time $t_0$ such that the distribution
$W[\varphi_0,\pi_0]$ is easy to compute.

Note that random elements of the Gaussian ensemble defined by
eq.~(\ref{eq:init-corr}) can be generated by writing
\begin{eqnarray}
  \varphi_0(\x)
  &=&
  \int \frac{d^3\q}{(2\pi)^3 2E_\q}\;
  \big[
  c_\q\,\varphi_{\q}(t_0,\x)
  +
  d_\q\,\varphi_{\q}^*(t_0,\x)
  \big]
  \nonumber\\
  \pi_0(\x)
  &=&
  \int \frac{d^3\q}{(2\pi)^3 2E_\q}\;
  \big[
  c_\q\,\dot\varphi_{\q}(t_0,\x)
  +
  d_\q\,\dot\varphi_{\q}^*(t_0,\x)
  \big]
  \; ,
  \label{eq:init0}
\end{eqnarray}
where $c_\q$ and $d_\q$ are random complex Gaussian distributed
numbers, whose 2-point correlations are
\begin{eqnarray}
  &&
  \big<c_\q c_{\q'}^*\big>= \big<d_\q d_{\q'}^*\big>=(2\pi)^3 E_\q \delta(\q-\q')
  \nonumber\\
  &&
  \big<c_\q c_{\q'}\big>= \big<d_\q d_{\q'}\big>=\big<c_\q d_{\q'}\big>=\big<c_\q d_{\q'}^*\big>=0\; .
\end{eqnarray}

\subsection{Numerical cost}
Eq.~(\ref{eq:LO2}) may at first appear to be a drawback compared to
eq.~(\ref{eq:LO1}) since we have replaced an ordinary integral by a
functional integration. However, let us assume that the mode function
$\varphi_{\q}$ is known (or at least easily computable) at the
initial time $t_0$. This is for instance the case when the background
field ${\cal A}^\mu$ is vanishing for $x^0<t_0$. Then, one can
estimate the functional integral by doing a Monte-Carlo sampling of
the Gaussian ensemble defined by eqs.~(\ref{eq:init-corr}), i.e. by
generating random functions of the form (\ref{eq:init0}) and by
solving the equation of motion for each of these samples. The
computational time of this method would be of the order of
\begin{equation}
N_{\rm samples} \times N_{\rm latt} \times \big(N_{\rm latt} + N_t\big)\; ,
\label{eq:t2}
\end{equation}
where $N_{\rm samples}$ is the number of samples used in the
Monte-Carlo evaluation of the functional integral. Note that the first
term, proportional to $N_{\rm latt}^2$, is due to the fact that in
eq.~(\ref{eq:init0}) there is a sum over $\q$ at each position
$\x$. This has to be repeated for each sample, but needs to be done
only at the initial time $t_0$. This Monte-Carlo method of evaluating
the particle spectrum is less costly than the direct method provided
that
\begin{equation}
N_{\rm samples}\ll N_{\rm latt}\quad,\quad N_{\rm samples}\ll N_t\; .
\end{equation}
The first condition is obvious: if one uses a number of samples which
is larger than the number of independent mode functions, then one
would be better off using the direct method (since it would give the
exact leading order answer, for a lesser computational effort). The
second condition implies that the computation is dominated by the
resolution of the equations of motion rather than the evaluation of
the initial conditions.

\section{Lattice numerical evaluation}
\label{sec:num-1loop}
\subsection{Lattice setup}
In the two formulations, (\ref{eq:LO1}) or (\ref{eq:LO2}), one needs
to solve the linearized equation of motion $({\cal D}_\mu{\cal D}^\mu+m^2)
\varphi=0$ for the propagation of a scalar field on top of some
background electromagnetic potential ${\cal A}_\mu$. There are only a
few known examples of background fields for which this equation of
motion can be solved analytically. For a generic background field, one
can only solve this equation numerically. 

Actual computations are done by discretizing space on a lattice.  We
consider a finite box of volume $L_x \times L_y \times L_z$ and divide
it into a $N_{\rm latt}\equiv N_x \times N_y \times N_z$ lattice.
Space points are labeled as
\begin{equation}
  \x \equiv (n_x a_x ,n_y a_y ,n_z a_x ) \quad 
  (n_x = 1,\cdots ,N_x ;\ n_y = 1,\cdots ,N_y ;\ n_z = 1,\cdots ,N_z )\; ,
\end{equation}
with lattice spacing $a_x =L_x /N_x$, etc. We impose the periodic
boundary conditions, e.g. $\varphi(x,y,z)=\varphi(x+L_x,y,z)$, and the
momenta are given by
\begin{equation}
p_i = \frac{2\pi k_i}{L_i} \hspace{20pt} (i=x,y,z)\; ,
\end{equation}
where the $k_i$ are integers, taken in the range 
\begin{equation}
k_i = -\frac{N_i}{2}+1,\cdots ,0,\cdots ,\frac{N_i}{2} \qquad (i=x,y,z)\; ,
\end{equation} 
where we have assumed that the lattice sizes $N_i$ are even. 

On the lattice, differentiation with respect to space is replaced by
finite differences.  Let us introduce the forward difference
\begin{equation}
\nabla _i ^+ \varphi (\x ) \equiv
 \frac{1}{a_i } \left[ \varphi(\x +a_i \hat{n}_i ) -\varphi(\x ) \right]\; ,
\end{equation}
where the vector $\hat{n}_i$ is the displacement by one lattice
spacing in the spatial direction $i$. Similarly, the backward difference is
\begin{equation}
\nabla _i ^- \varphi (\x ) \equiv \frac{1}{a_i } \left[ \varphi(\x ) -\varphi(\x -a_i \hat{n}_i ) \right]\; .
\end{equation}
By ``integration by parts'', the forward difference is transformed to
the backward difference~:
\begin{eqnarray}
\sum _\x \big[\nabla_i ^+ f(\x )\big]g(\x ) 
 =
 -\sum _\x f(\x ) \Big[\nabla_i ^- g(\x )\Big]\; . 
\end{eqnarray}
(Notice that there is no boundary term because of the periodic boundary
condition.) In words, this means that $\nabla _i ^-$ and $\nabla _i
^+$ are mutually adjoint. Since it is desirable to have a self-adjoint
Laplacian operator, it is convenient to define it by a mix of forward
and backward derivatives
\begin{equation}
\Delta \equiv \sum_{i=x,y,z} \nabla_i ^- \nabla_i ^+\; . 
\end{equation}
The discrete plane waves $\exp(i\p\cdot\x)$ are eigenfunctions of this
operator, with eigenvalues
\begin{equation}
-E_{k_x,k_y,k_z}^2\equiv -2\sum_{i=x,y,z}
\frac{\sin^2(\frac{\pi k_i}{N_i})}{a_i^2}
\; .
\end{equation}

When considering scalar QED, the local  $U(1)$ gauge invariance can be
preserved on the lattice by defining the forward covariant derivative as
\begin{equation}
D_i ^+ \varphi (\x ) \equiv \frac{1}{a_i } \left[ e^{ie a_i A_i (\x )} \varphi (\x +a_i \hat{n}_i ) -\varphi (\x ) \right]\; ,
\end{equation} 
Under a gauge transformation\footnote{There is some arbitrariness in
  how we discretize the gauge transformation law for $A_i$, since we
  could have chosen a backward derivative $\nabla_i^-$ instead of the
  forward derivative. If we adopt this alternative choice, the forward
  covariant derivative must be changed into
  \begin{equation*}
    D_i ^+ \varphi (\x ) \equiv \frac{1}{a_i } \left[ e^{ie a_i A_i (\x+ a_i \hat{n}_i)} \varphi (\x +a_i \hat{n}_i ) -\varphi (\x ) \right]\; .
  \end{equation*}}
\begin{eqnarray}
\varphi (t,\x ) &\longrightarrow&
 e^{ie\theta (t,\x )} \varphi (t,\x ) \nonumber\\
A_0 (t,\x ) &\longrightarrow&
 A_0 (t,\x ) -\dot{\theta} (t,\x ) \nonumber\\
A_i (t,\x ) &\longrightarrow& 
A_i (t,\x ) -\nabla _i ^+ \theta (t,\x ) \; , 
\label{eq:GT}
\end{eqnarray}
$D_i ^+ \varphi (t, \x )$ is transformed in the same way as $\varphi (t,\x )$~: 
\begin{equation}
D_i ^+ \varphi (t, \x ) \;\longrightarrow\; e^{ie\theta (t,\x )} D_i ^+ \varphi (t, \x )\; . 
\end{equation}
One can write a gauge invariant discretized Lagrangian density for the
complex scalar fields as follows
\begin{equation} \label{eq:Lmatter}
\mathcal{L}_\text{matter} 
 = 
\left( D_0 \phi \right)^* \left(D_0 \phi\right)
 -\sum_{i=x,y,z} \left( D_i ^+ \phi \right)^* \left(D_i ^+ \phi\right)
 -m^2\phi^*\phi-V\left( \phi^* \phi \right)\; . 
\end{equation}
In deriving the discretized classical equation of motion, one should note that
the forward covariant derivative is the adjoint of the backward
covariant derivative $D_i^-$, 
\begin{equation}
D_i ^- \varphi (\x ) = \frac{1}{a_i } \left[ \varphi (\x ) -e^{-ie a_i A_i (\x -a_i \hat{n}_i )} \varphi (\x -a_i \hat{n}_i ) \right]\; . 
\end{equation}
One obtains
\begin{equation}
\Big(D_0^2-\sum_{i=x,y,z}D_i^- D_i^++m^2\Big)\,\varphi+V'(\varphi^*\varphi)\,\varphi=0\; .
\end{equation}
(Here the equation is written with the self-interaction term, but in
the evaluation of the spectrum at leading order we need only the
linear part of the equation.) 
It is convenient to choose the temporal gauge $A^0=0$, so that
$D_0=\partial_0$ in the equation of motion.

\subsection{Numerical results}
In order to demonstrate that the classical statistical simulation
(CSS) can indeed describe the Schwinger mechanism at leading order, we
consider a simple situation which can be handled easily by the direct
quantum field theory method. The self-coupling $\lambda$ is set to
zero in this leading order computation, and we take a spatially
homogeneous background electrical field in the $z$ direction (and no
magnetic field), that we switch on at the time $x^0=0$. In this case,
thanks to the translational invariance of the problem, the direct
evaluation of eq.~(\ref{eq:LO1}) is not very expensive and will be
used as a benchmark against which we compare the CSS results.

The parameters of the lattice simulation are $N_x=N_y=32$, $N_z=256$
(we need more lattice spacings in the direction of the electrical
field), corresponding to physical sizes $L_x=L_y=50$, and $L_z=30$
respectively. The mass is taken to be $m=0.1$, the electrical charge
is $e=1$ and the electrical field is switched on\footnote{In the
  temporal gauge $A^0=0$, this can be realized by the following gauge
  potential~:
\begin{eqnarray*}
 {\cal A}^1={\cal A}^2=0\quad,\quad -\partial^0 {\cal A}^3 = E\theta(x^0)\; ,
\end{eqnarray*}
which can be realized by the external current $J^\mu_{\rm ext} =
-\delta^{\mu 3}E \delta(x^0)$.} at $x^0=0$ and its value is $E=1$
(alternatively, one may view $eE$ as arbitrary and all the other
dimensionful quantities as being quoted in units of $\sqrt{eE}$, that
has the dimension of a mass). 1024 field configurations were used in
order to sample the Gaussian ensemble defined in
eqs.~(\ref{eq:init0}).

\begin{figure}[htbp]
\begin{center}
\resizebox*{7.5cm}{!}{\includegraphics{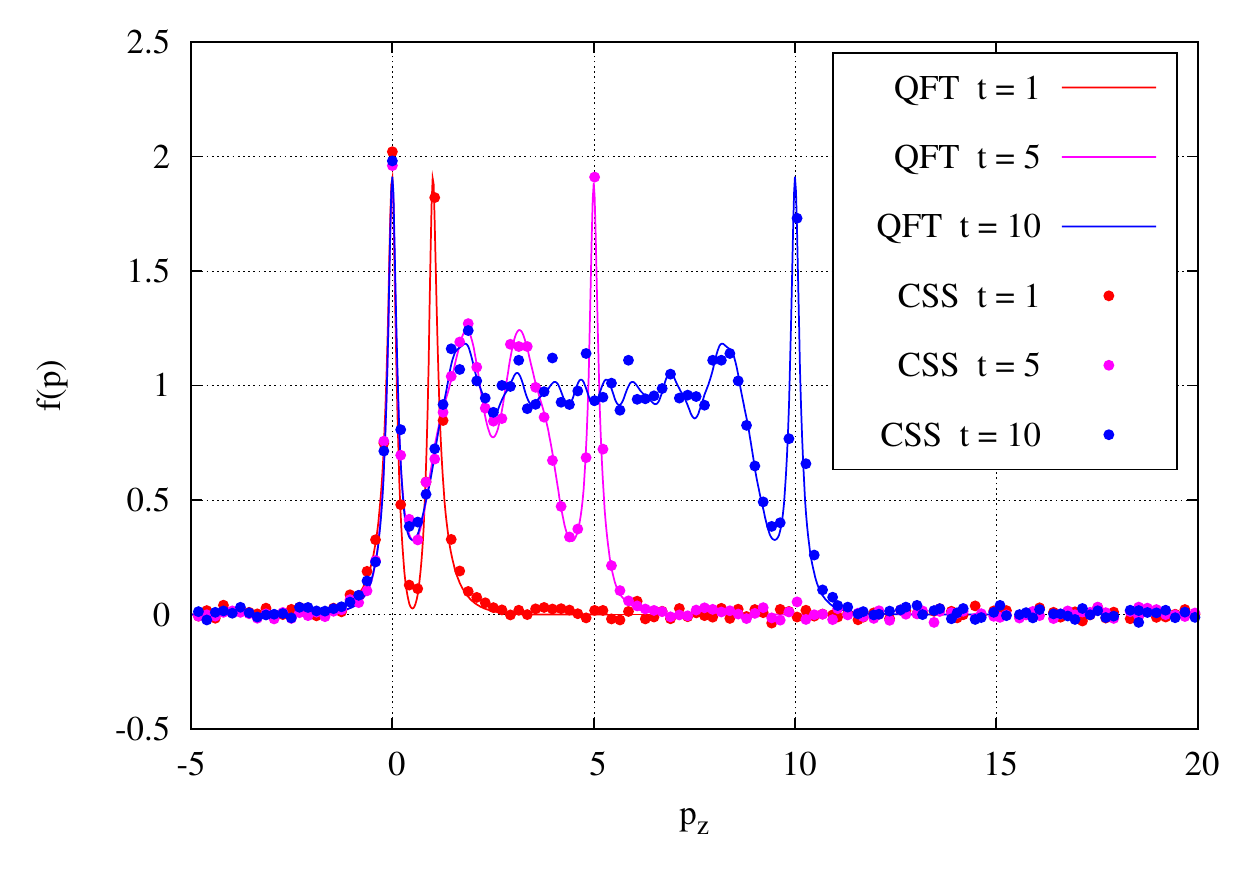}}
\end{center}
\caption{\label{fig:compCSS}Comparison of the $p_z$ spectrum between
  the classical statistical simulation and the direct 1-loop QFT
  calculation in the case of a constant electrical field.}
\end{figure}

In the figure \ref{fig:compCSS} we show the longitudinal momentum
distribution\footnote{The occupation number $f(\p)$ differs from the
  particle spectrum defined in eq.~(\ref{eq:LO2}) by a factor of the
  volume, $dN_1/d^3\p \equiv V\,f(\p)/(2\pi)^3$.} of the produced scalar
particles, at several times shortly after the electrical field has
been switched on. The dots represent the result of the classical
statistical approach and the solid lines are the direct QFT
calculation.
The agreement between the two approaches is very good, and the
differences compatible with the expected statistical error given the
number of samples used in the CSS approach. In particular, the
intricate oscillatory pattern of this spectrum, which results from
quantum interference phenomena, is well reproduced in the CSS method.
  This shows very concretely how 1-loop quantum effects can be
  reformulated in terms of purely classical objects.

\begin{figure}[htbp]
  \begin{center}
  \resizebox*{7.5cm}{!}{\includegraphics{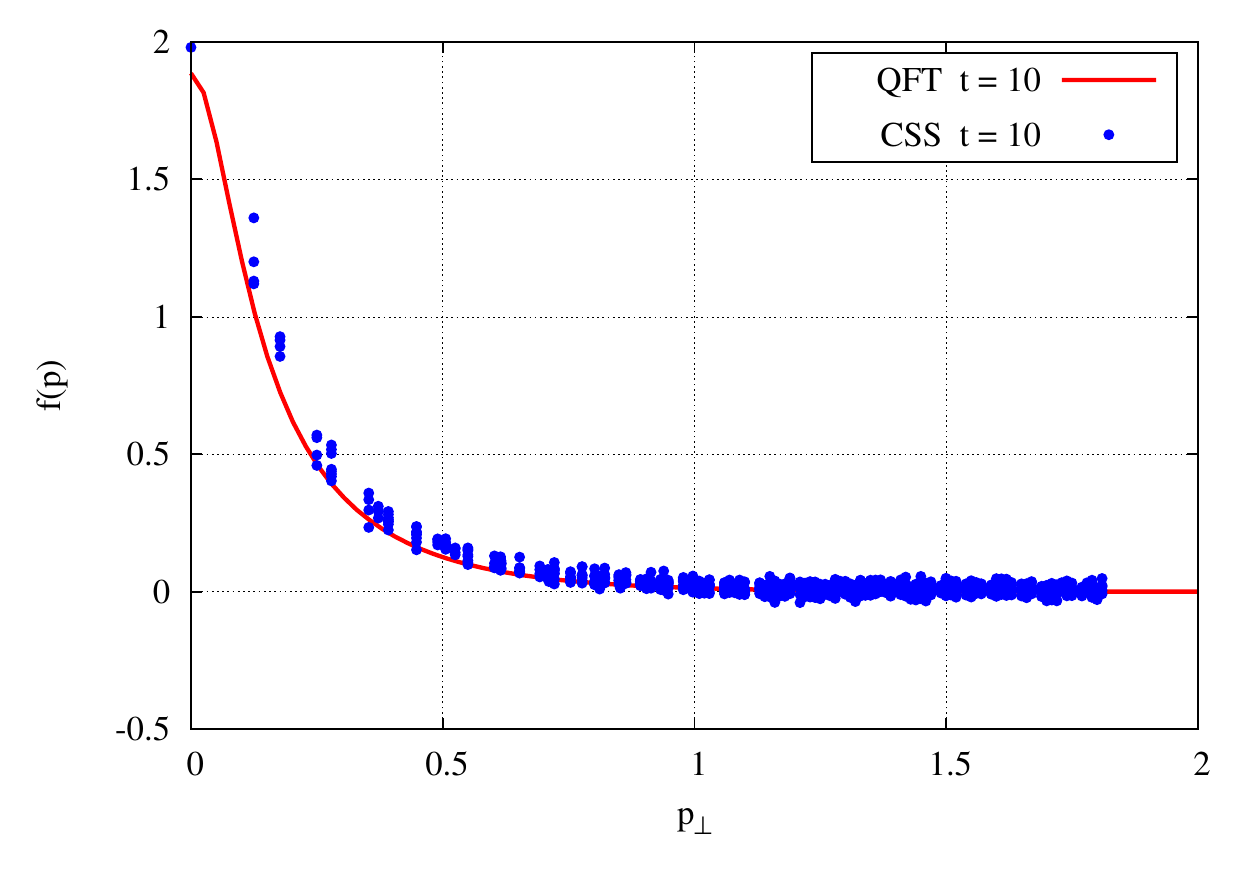}}
  \end{center}
  \caption{\label{fig:compCSS-pt}Comparison of the $p_\perp$ spectrum
    between the classical statistical simulation and the direct 1-loop
    QFT calculation in the case of a constant electrical field.}
  \end{figure}

  The time evolution of the $p_z$ spectrum is rather transparent:
  particles are produced with a small $p_z$ by quantum tunneling, and
  later they are accelerated in the direction of the electrical field;
  hence the expansion of the spectrum towards larger (positive, because
  we are considering only particles --not antiparticles--) values of
  $p_z$. One can indeed see on the figure \ref{fig:compCSS} that this
  expansion is linear in time, in good agreement with a constant
  acceleration $eE$ (which is equal to 1 with our choice of units) of
  the particles in the $+z$ direction. Similarly, the comparison of the
  $p_\perp$ spectra obtained in the two approaches, in the figure
  \ref{fig:compCSS-pt}, show a good agreement within statistical errors.
  The shape of the transverse momentum spectrum is very different from
  that of the longitudinal momentum distribution. Roughly speaking, the
  produced particles originate from virtual pairs (i.e. vacuum
  fluctuations) that can have a momentum in any direction. Then, their
  longitudinal momentum $p_z$ increases linearly in time due to the
  electrical field, while their transverse momentum is not affected.

\begin{figure}[htbp]
  \begin{center}
  \resizebox*{6cm}{!}{\includegraphics{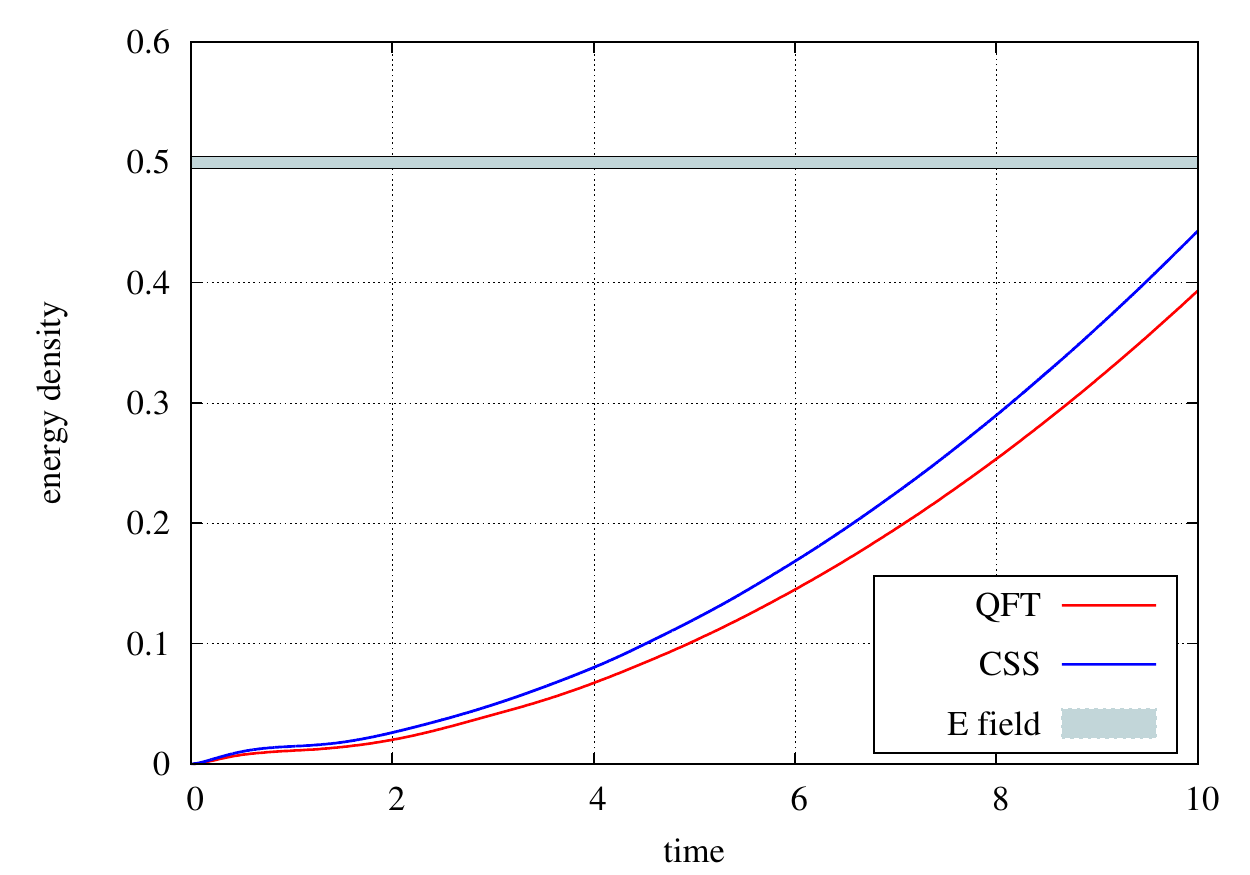}}
  \hfil
  \resizebox*{6cm}{!}{\includegraphics{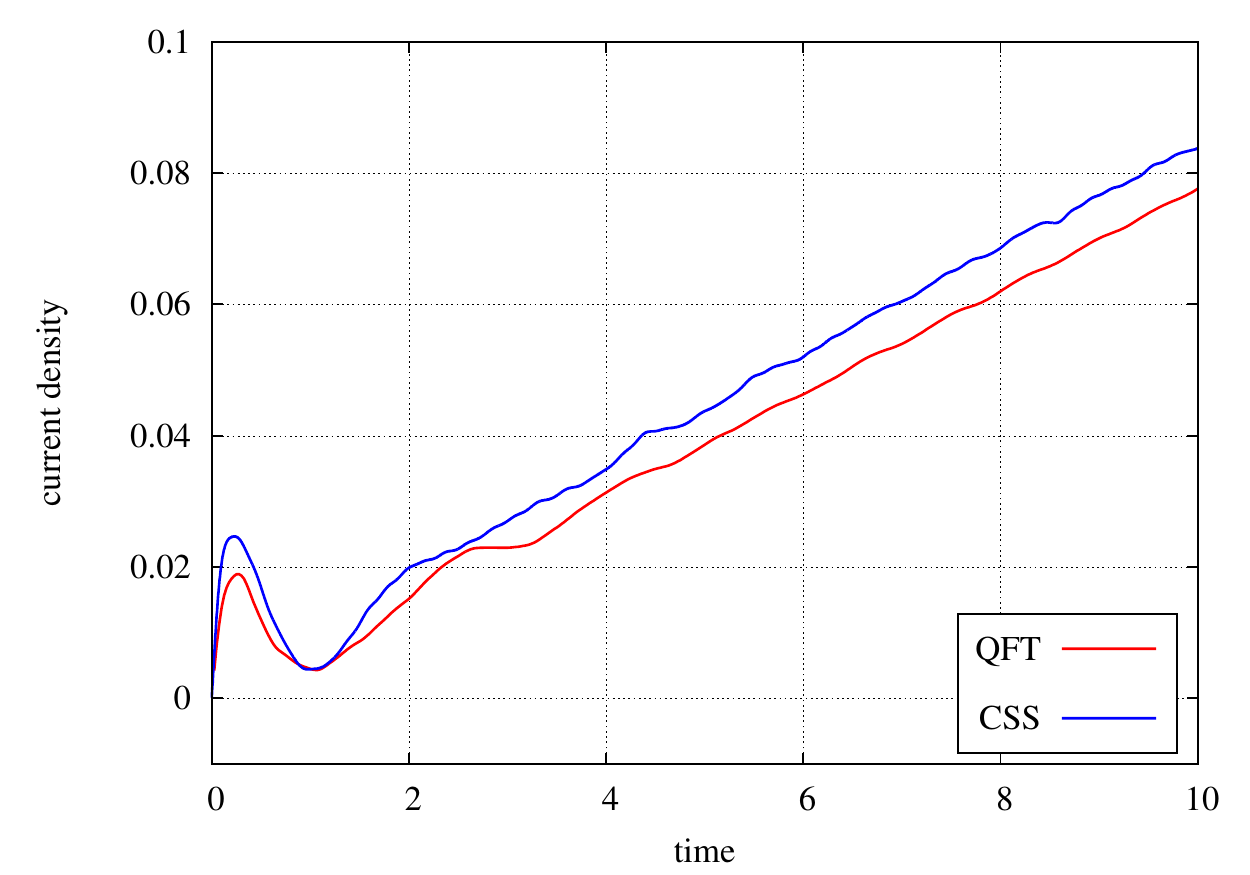}}
  \end{center}
  \caption{\label{fig:en-cur}Energy density and electrical current
    density carried by the produced particles. In the left plot, the
    light blue band indicates the energy density of the electrical
    field.}
  \end{figure}

  To close this section, we also show the energy density and electrical
  current density carried by the produced particles, as a function of
  time, in the figure \ref{fig:en-cur}.
  The energy density of the produced particles increases steadily with
  time, and at some point it will overcome the energy density of the
  background electrical field. When this occurs, the simple calculation
  done in this section, where the back-reaction of the produced
  particles on the gauge field is neglected, becomes certainly
  insufficient. In particular, the total energy present in the system is
  not conserved at this level of approximation, because of the
  uninterrupted production of scalar particles, shown in the left plot
  of the figure \ref{fig:en-cur}. One could in fact use this as a
  criterion for deciding when the one loop approximation ceases to be
  valid: it should be improved when the energy carried by the produced
  particles becomes of the same order of magnitude as the energy density
  carried by the electrical field. From the left plot of
  \ref{fig:en-cur}, we expect this breakdown to occur as early as
  $t\sim10$.\;\footnote{%
  Time is scaled like $1/\sqrt{eE}$. 
  For the QED critical field strength $E=m^2/e$, $t\sim 10$
  amounts to $t\sim 10^{-20}$ seconds. 
  }  
  This is also corroborated by the behavior of the electrical
  current carried by the produced particles. Because this current acts
  as a source in the Maxwell's equations that control the gauge
  potential, it can alter the background electrical field when it
  becomes too large.

  \section{Back-reaction effects}
  \label{sec:br}
  In the previous section, we have seen that the plain 1-loop
  calculation of the spectrum of produced particles is bound to break
  down after some time, because this approximation violates energy
  conservation. What is missing is the back-reaction of the produced
  scalar particles on the electrical field, which has the effect of
  screening this field, which eventually will put an end to the
  production of particles~\cite{KlugeESCM1,GavriG1}.

  Taking into account the back-reaction means that the source of the
  electromagnetic field is not just the external source $J^\mu_{\rm
    ext}$, but also includes the contribution of the scalar field $\phi$
  to the electromagnetic current,
  \begin{eqnarray} \label{eq:Jind} 
  j^0(x) &=&
   ie \left[ \phi ^* (x) \dot{\phi} (x) -\dot{\phi} ^* (x) \phi (x) \right]
  \nonumber\\
  j^i (x) &=&
   ie \left[ \phi^* (x) D_i^+ \phi (x) -\left( D_i^+ \phi (x) \right)^* \phi (x) \right]\; ,
  \end{eqnarray}
  written here with lattice discrete spatial derivatives. In the gauge
  $A^0=0$, the Maxwell's equation for the vector potential read
  \begin{eqnarray}
  &&\nabla^i_- \dot{A}^i = J^0_{\rm ext}+j^0\nonumber\\
  &&\ddot{A}^i +(\nabla^i_+\nabla^j_--\delta^{ij}\nabla_+\cdot\nabla_-)\,A^j = J^i_{\rm ext}+j^i\; .
  \label{eq:Maxwell}
  \end{eqnarray}
  The first of the two equations (\ref{eq:Maxwell}) is Gauss's law. It
  is automatically satisfied if $A^i$ obeys the second equation, and
  $J^\mu$ is conserved\footnote{This follows from the identity
    $\nabla^i_-(\nabla_+\cdot\nabla_-)=(\nabla_-\cdot\nabla_+)\nabla^i_-$.},
  \begin{equation}
  \dot{J}^0=\nabla^i_- J^i\; ,
  \end{equation}
  which, in turn, is the case if the scalar field $\phi$ obeys the
  equation of motion.
  \begin{figure}[htbp]
  \begin{center}
    \resizebox*{3.5cm}{!}{\includegraphics{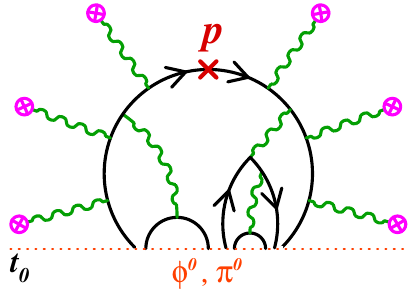}}
    \hfil
    \resizebox*{3.5cm}{!}{\includegraphics{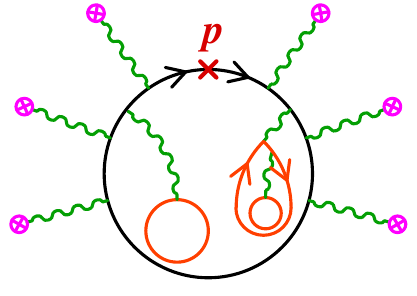}}
  \end{center}
  \caption{\label{fig:br}Left: topologies included when solving
    simultaneously the equations of motion of the scalar field $\phi$
    and the Maxwell's equations with the induced current, from some
    given boundary conditions $\varphi_0,\pi_0$ at the initial time
    $t_0$. Right: QFT topologies when we perform a Gaussian average of
    the left graph over the Gaussian ensemble (\ref{eq:init-corr}), with
    $j^\mu$ replaced by the ensemble average $\big<j^\mu\big>$ in
    Maxwell's equations.}
  \end{figure}
  Diagrammatically, solving simultaneously the equation of motion of the
  scalar field $\phi$ and Maxwell's equations, starting from some
  initial condition $\varphi_0(\x),\pi_0(\x)$ at the time $t_0$, amounts
  to resumming graphs such as the one represented in the left part of
  the figure \ref{fig:br}. All the graphs are made of a principal scalar
  line, to which the measured particle of momentum $\p$ is attached. To
  this line are attached a number of photons. These photons can either
  be attached to the external current $J^\mu_{\rm ext}$, or to the
  induced current, i.e. to another scalar line. These secondary scalar
  lines can themselves be decorated by photons, etc...

  Averaging the functional of $\varphi_0,\pi_0$ obtained by this
  procedure over the Gaussian distribution of initial conditions defined
  by (\ref{eq:init-corr}) amounts to reconnecting pairwise all the
  hanging scalar lines in the graph of the figure \ref{fig:br}.
  Although all the topologies obtained when doing this were indeed
  present in the original quantum field theoretical formulation of the
  particle spectrum, some of them are miscalculated by the classical
  statistical simulation. This because in the CSS all propagators are
  retarded, while they are Feynman propagators in the field theoretical
  calculation\footnote{In other words, the CSS and the original quantum
    field theory differ by some commutators, as one may expect.}.
  However, when we reconnect together the two scalars that enter in the
  same instance of the induced current, the classical statistical
  approach reproduces exactly the QFT value of that loop. This amounts to
  replacing the induced current $j^\mu$ in Maxwell's equation by its
  ensemble average,
  \begin{equation}
  j^\mu\quad\to\quad \left<j^\mu\right>\; .
  \end{equation}
  By doing this, one obtains all the graphs such as the one represented
  in the right part of the figure \ref{fig:br}, where the scalar loops
  represented in orange originate from the use of $ \left<j^\mu\right>$
  in the right hand side of Maxwell's equations. Although this is not
  manifest on this example, these scalar loops can themselves be dressed
  by an arbitrary number of photon insertions, whose other endpoint can
  either be the external current $J^\mu_{\rm ext}$ or another instance
  of the ensemble averaged induced source.

\begin{figure}[htbp]
  \begin{center}
  \resizebox*{6cm}{!}{\includegraphics{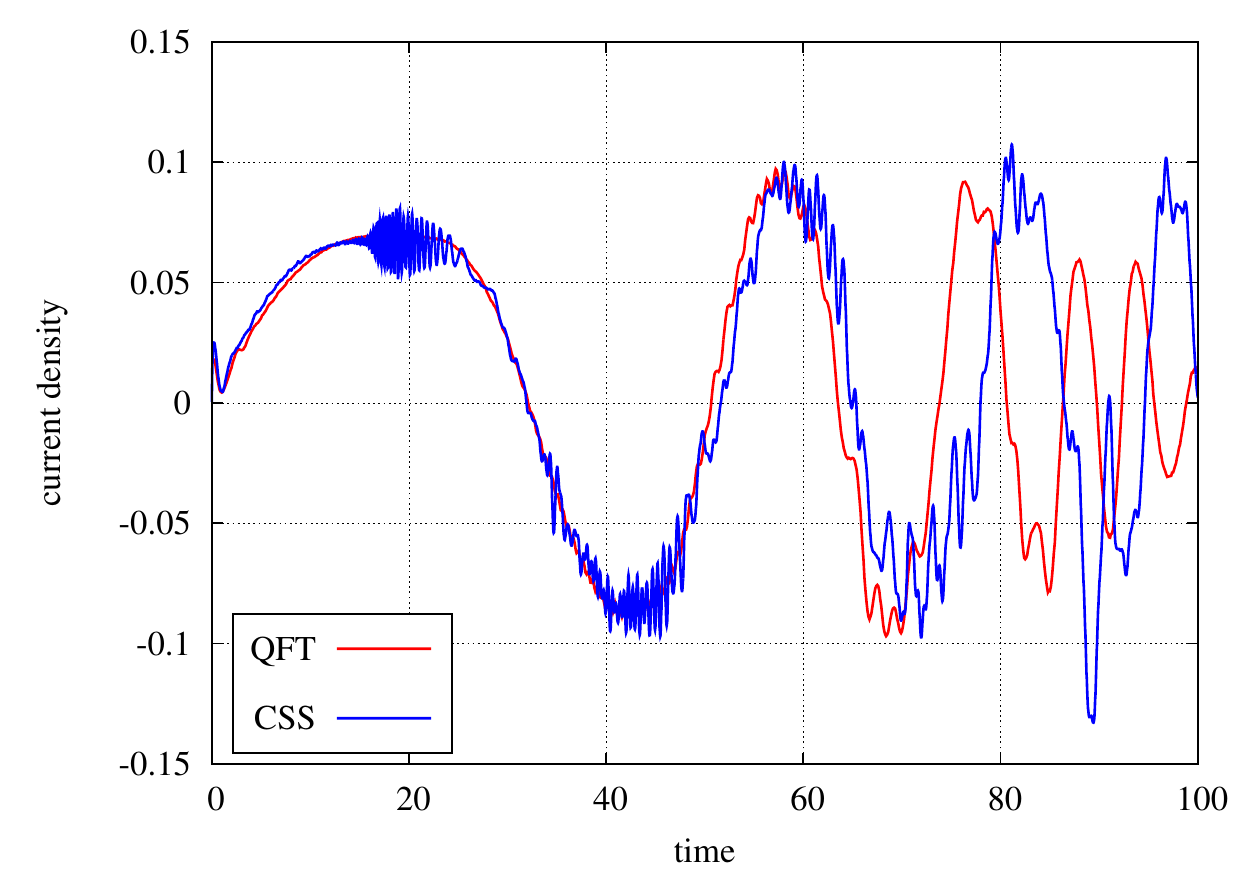}}
  \hfil
  \resizebox*{6cm}{!}{\includegraphics{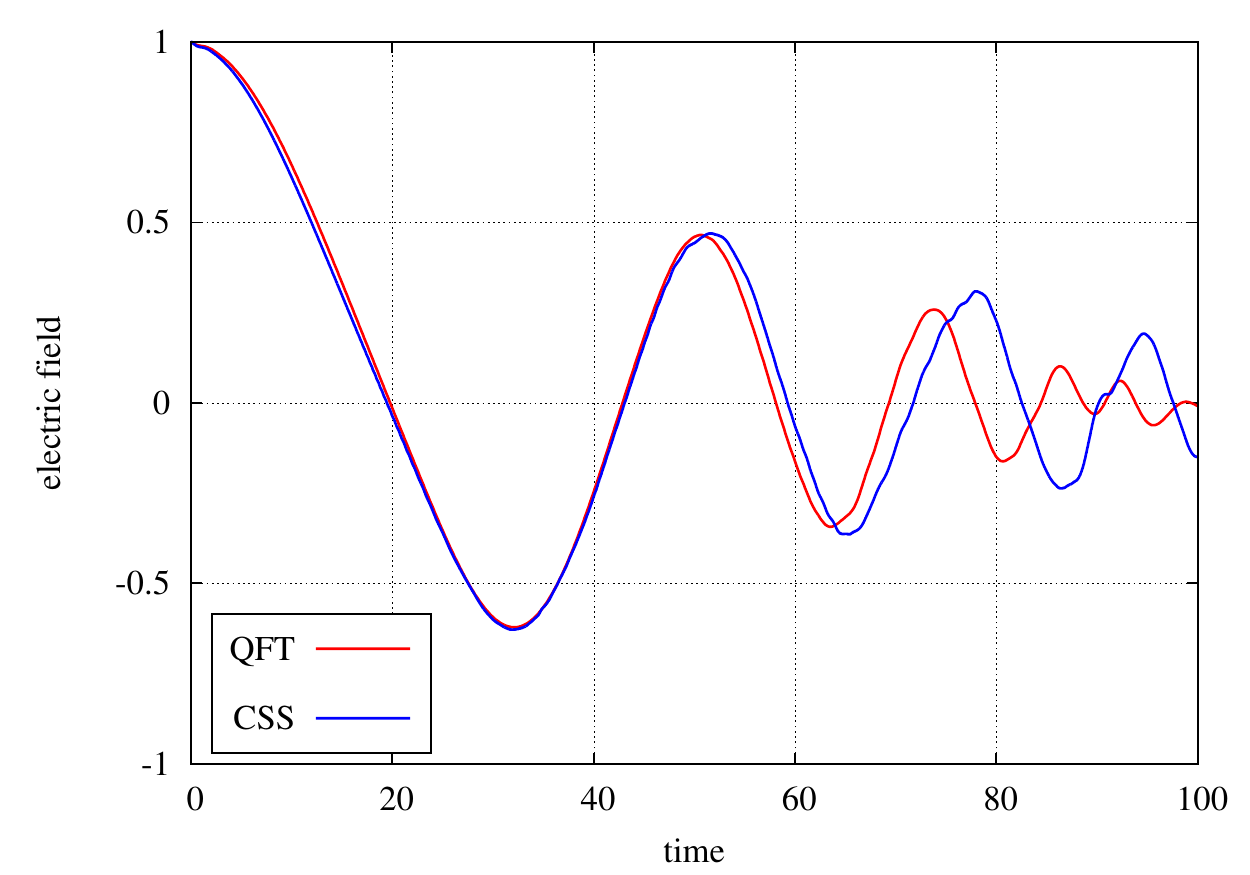}}
  \end{center}
  \caption{\label{fig:br1}Time evolution of the electrical current
    density and of the resulting electrical field when the back-reaction
    is taken into account.}
  \end{figure}

  Let us now show some numerical results that illustrate the effect of
  the back reaction. First, in the figure \ref{fig:br1}, we display the
  electrical current density (left) and the resulting electrical field
  (right).
  There in an inflection in the growth of the current at a time $t\approx
  10$, which roughly corresponds to the moment when the energy carried
  by the produced particles becomes comparable to the energy stored in
  the electrical field. Simultaneously, the electrical field decreases
  and even changes sign periodically, while the amplitude of its
  oscillations decrease to zero. Consequently, the production of
  particles slows down and effectively stops after some time (when the
  probability of particle creation, of the order of $\exp(-\pi m^2/ eE)$,
  becomes too small).
  \begin{figure}[htbp]
  \begin{center}
  \resizebox*{4cm}{!}{\includegraphics{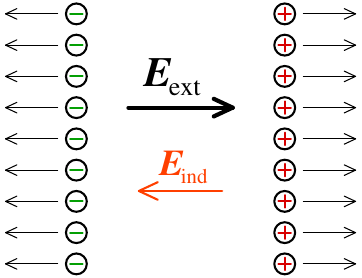}}
  \end{center}
  \caption{\label{fig:ind}Illustration of the polarization phenomenon
    that leads to the reduction of the electrical field when the
    back-reaction is taken into account.}
  \end{figure}
  It is rather easy to understand why the induced electrical field is
  opposite in direction to the externally field, as illustrated in the
  figure \ref{fig:ind}. Indeed, after having been produced, the positive
  charges are accelerated in the direction of the external field and the
  negative charges in the opposite direction. This charge separation
  acts like capacitor plates, that create an induced field oriented from
  the positive to the negative charges\footnote{%
  In a semiclassical tunneling picture, 
  a particle and an antiparticle are produced with a distance 
  $\frac{2m^2}{eE}$ between them. 
  This initial separation contributes to the polarization current, 
  while the subsequent charge acceleration does the conduction current \cite{Tanji1}. 
  }. 
  The induced electrical field
  thus counteracts the effect of the external field.

\begin{figure}[htbp]
  \begin{center}
  \resizebox*{6cm}{!}{\includegraphics{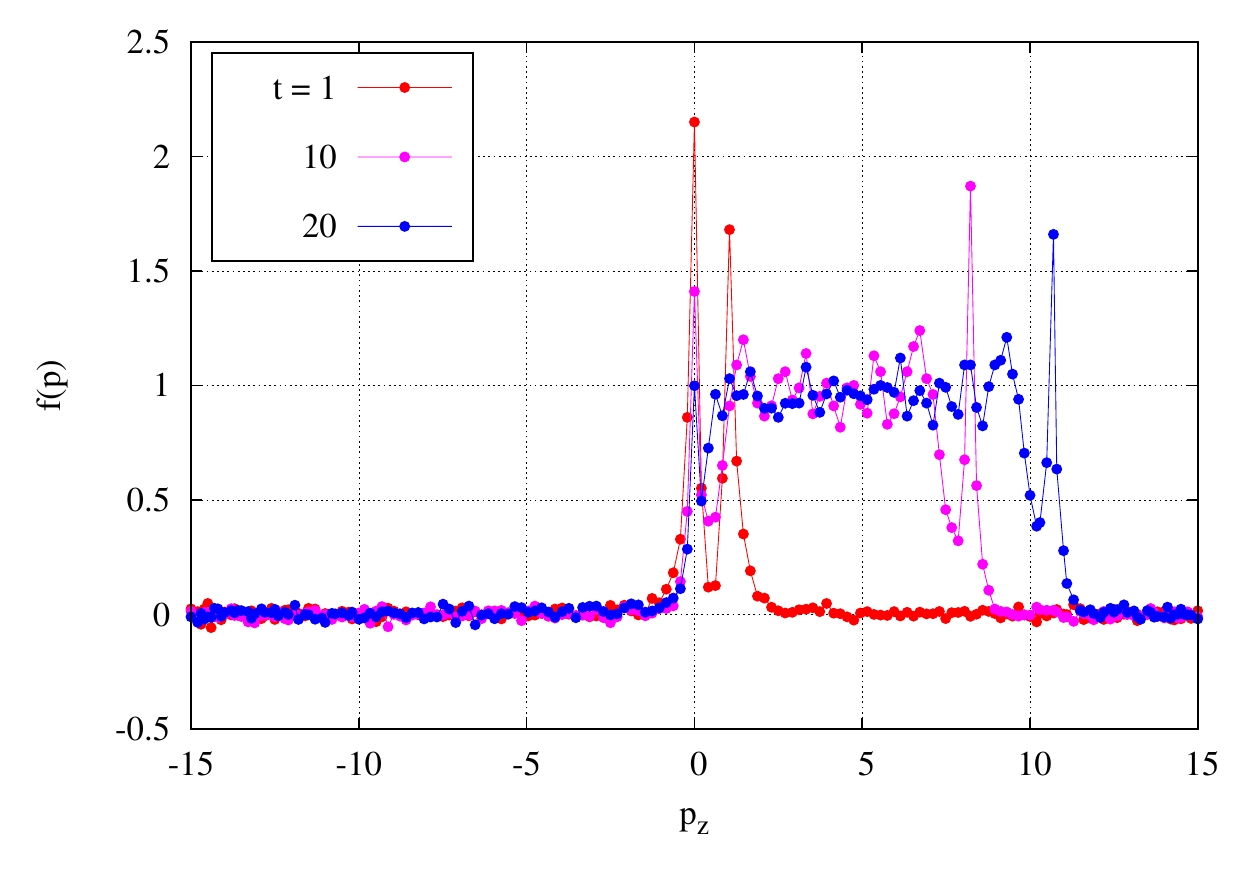}}
  \hfil
  \resizebox*{6cm}{!}{\includegraphics{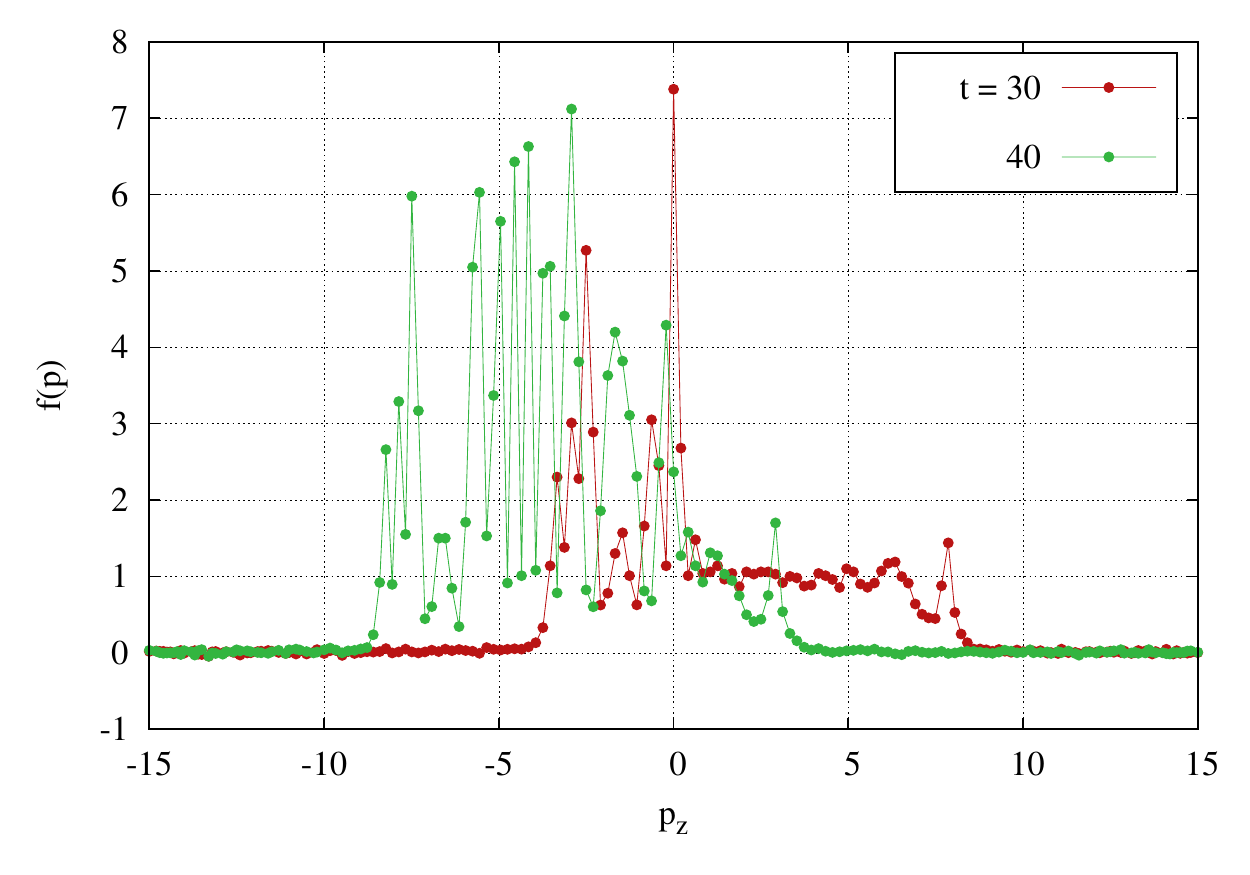}}
  \end{center}
  \caption{\label{fig:br2}Left: evolution at short times of the $p_z$
    spectrum of the produced scalar particles. Right: evolution of the
    $p_z$ spectrum at intermediate times.}
  \end{figure}

\begin{figure}[htbp]
  \begin{center}
  \resizebox*{6cm}{!}{\includegraphics{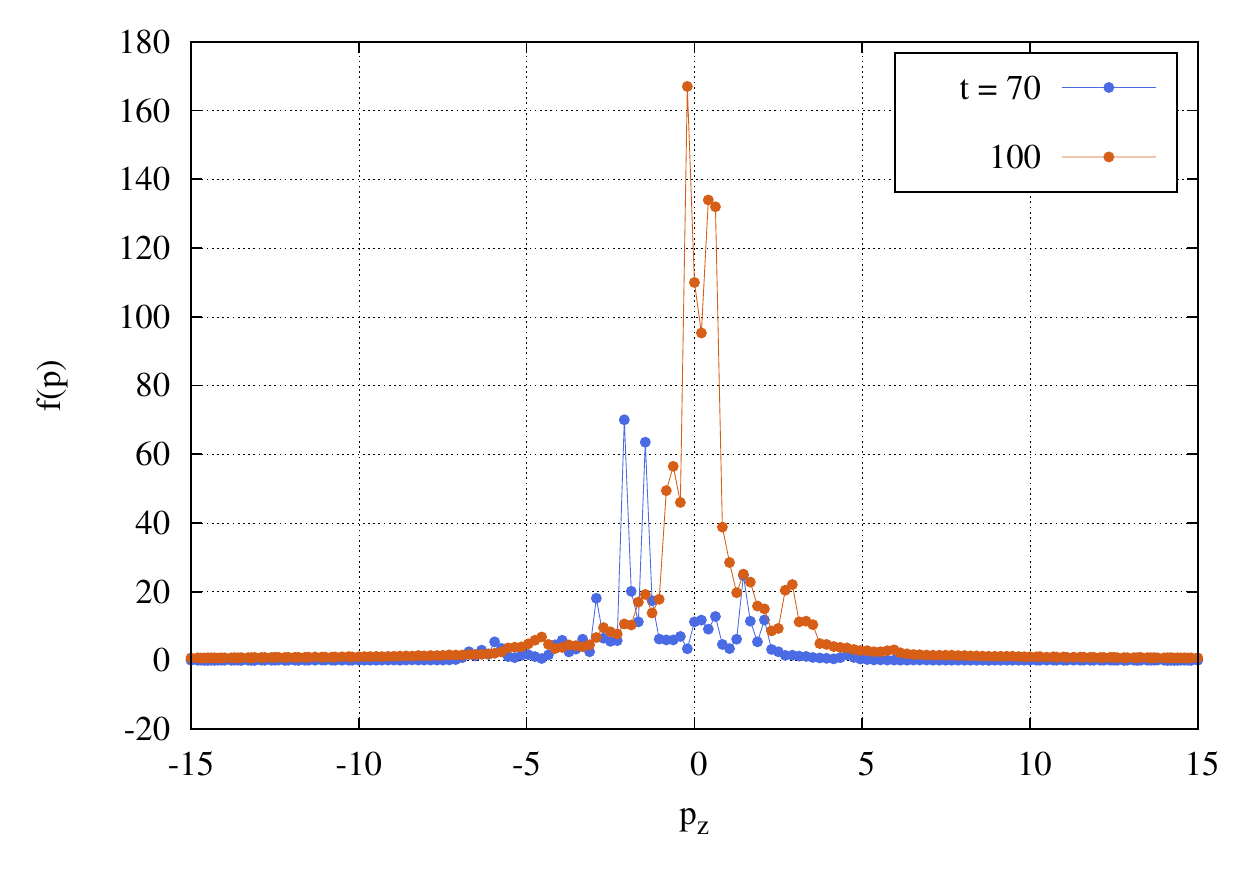}}
  \hfil
  \resizebox*{6cm}{!}{\includegraphics{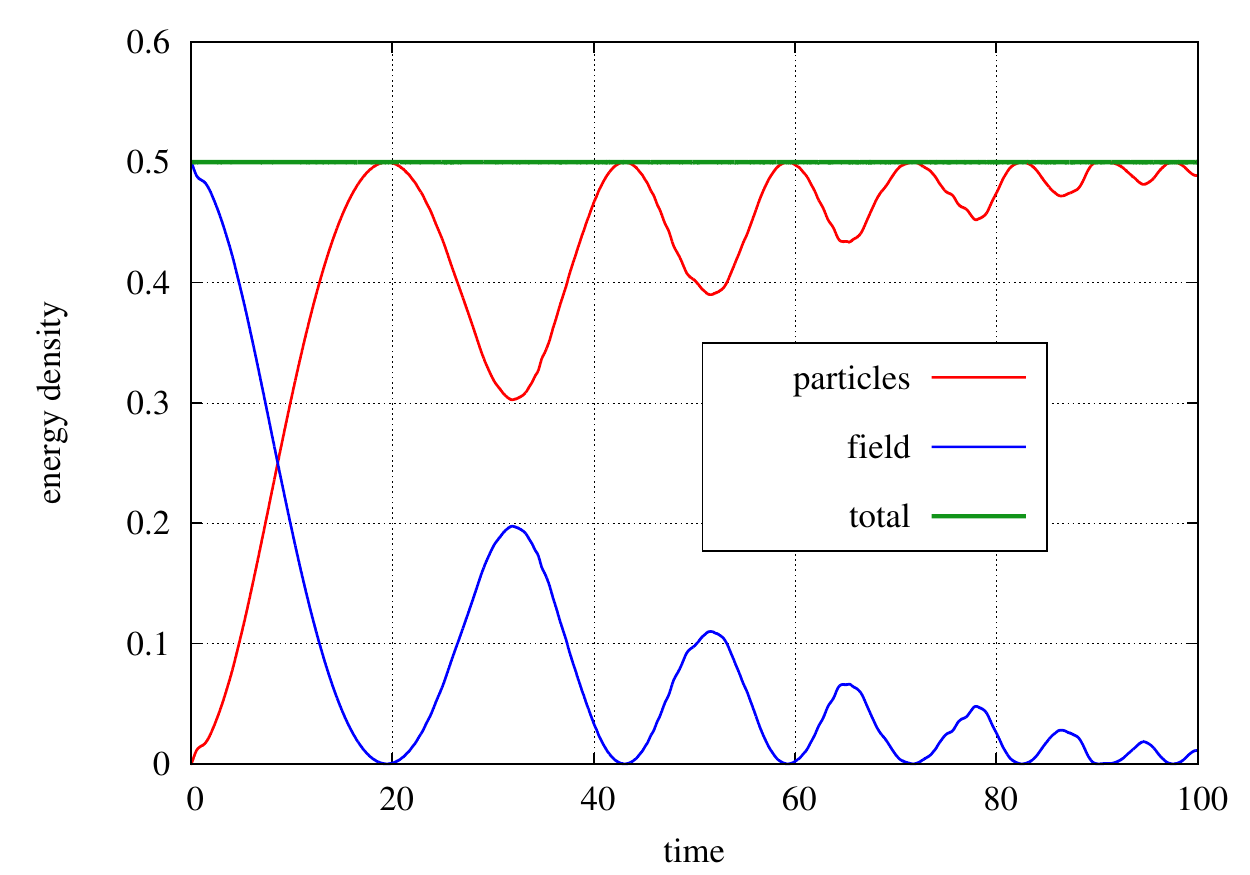}}
  \end{center}
  \caption{\label{fig:br3}Left: evolution of the $p_z$ spectrum over
    longer time scales. Right: decomposition of the energy of the system
    between fields and particles.}
  \end{figure}

  In the left plot of the figure \ref{fig:br2}, we see that the
  expansion of the $p_z$ spectrum towards the right is no longer linear
  in time, a direct consequence of the decreasing electrical field which
  is no longer providing a constant acceleration to the produced
  particles. In fact the plot on the right of the figure \ref{fig:br2}
  shows that for slightly larger times, the shift towards the right of
  the $p_z$ spectrum comes to a halt, and is replaced by a shift to the
  left. Obviously, this happens when the electrical field has changed
  sign, around $t\approx 30$.
  After the $p_z$ spectrum moves into the negative momentum region, its
  value undergoes a rapid increase. This is because particles which have
  been created earlier pass through the zero-momentum region and
  stimulate the subsequent particle production (Bose-enhancement).  At
  the same time, the $p_z$ spectrum starts to show an oscillatory
  pattern.  This is due to an interference between the fields of the
  previously produced particles and of the newly produced
  particles~\cite{Tanji2}.  It is remarkable that the CSS method can
  describe this intricate pattern of peaks, that are purely quantum
  effects.
  At even larger times, when the electrical field has become small, the
  $p_z$ distribution becomes roughly centered around $p_z=0$, as one can
  see from the plot on the left of the figure \ref{fig:br3}.
  Because the electric field has changed its sign for several times, the
  spectrum has gone through several stages of Bose-enhancement and is
  now much larger than its value at early times.
  Finally, one can also check that the inclusion of the back-reaction
  effects cure the main problem of the plain 1-loop calculation,
  i.e. the energy conservation. In the plot on the right of the figure
  \ref{fig:br3}, we have represented the energy density carried by the
  produced particles, the energy density carried by the electromagnetic
  fields, and the sum of the two contributions. We observe a transfer
  into particles of the energy initially stored in the electrical field,
  while the total energy remains constant.

  \section{Self-interactions and mass renormalization}
  \label{sec:self}
  So far, all the results we have shown have neglected the
  self-interactions of the scalar fields. This means that after the
  external electrical field has been ``neutralized'' by the produced
  particles, their distribution is frozen and ceases to evolve.  In
  particular, it has no way to thermalize.  The effect of these
  self-interactions has been studied in the classical statistical
  framework employed here, where it is rather straightforward to take
  into account since it just amounts to adding a non-linear term in the
  equation of motion of the classical scalar fields,
  \begin{equation}
  \Big(D_0^2-\sum_{i=x,y,z}D_i^- D_i^++m^2\Big)\,\varphi+\frac{\lambda}{2}(\varphi^*\varphi)\,\varphi=0\; ,
  \end{equation}
  written here for a quartic self-interaction. In the slightly simpler
  example of a real scalar field theory, it has been shown that these
  nonlinearities lead to the isotropization and
  thermalization\footnote{Since it is a semi-classical approximation,
    the asymptotic spectrum obtained in the classical statistical
    framework is not a full fledged Bose-Einstein distribution but only
    its soft part $f(\p)=T/E_\p$.}  of the momentum distribution of the
  particles.
  \begin{figure}[htbp]
  \begin{center}
    \resizebox*{3.5cm}{!}{\includegraphics{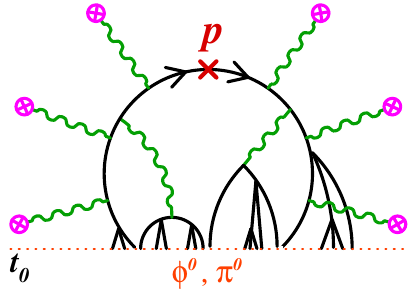}}
    \hfil
    \resizebox*{3.5cm}{!}{\includegraphics{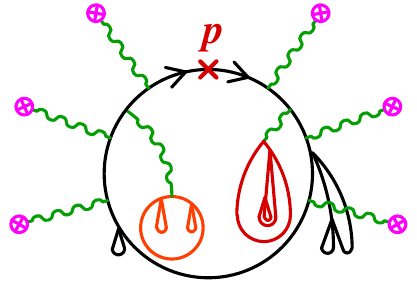}}
  \end{center}
  \caption{\label{fig:br-si}Left: topologies included when solving
    simultaneously the equations of motion of the scalar field $\phi$
    with self-interactions and the Maxwell's equations with the induced
    current, from some given boundary conditions $\varphi_0,\pi_0$ at
    the initial time $t_0$. Right: topologies obtained after the
    Gaussian average over the initial conditions $\varphi_0,\pi_0$, with
    $\big<j^\mu\big>$ as the source term in Maxwell's equations.}
  \end{figure}
  Diagrammatically, including the self-interactions in the classical
  equation of motion of the scalar fields changes the figure
  \ref{fig:br} into the figure \ref{fig:br-si}. In particular, the
  Gaussian average over the initial conditions for the scalar field can
  now produce loop corrections such as (but not only) tadpoles.

  Our goal in this section in not to reproduce previous results on
  thermalization in classical statistical simulations, but to stress the
  complication due to mass renormalization\footnote{One can find other
    discussions of renormalization in classical approaches in
    refs.~\cite{AartsS1,AartsS3}.}, which is crucial when dealing with a
  tunneling phenomenon such as the Schwinger mechanism. The main issue
  is that the probability of particle production by quantum tunneling is
  extremely sensitive to the value of the mass of the scalar particles,
  since its square enters in the exponential $\exp(-\pi
  m^2/eE)$. However, when we include the nonlinear term in the equation
  of motion of the scalar field and we average over its initial
  conditions, we resum some loop corrections that are ultraviolet
  divergent. In a lattice simulation, they are regularized by the
  lattice spacing, but provide a potentially large renormalization of
  the mass that will alter significantly the production of particles by
  the Schwinger mechanism. The worst offenders are the tadpole
  corrections, that have a quadratic dependence on the inverse lattice
  spacing.

  \begin{figure}[htbp]
  \begin{center}
  \resizebox*{8cm}{!}{\includegraphics{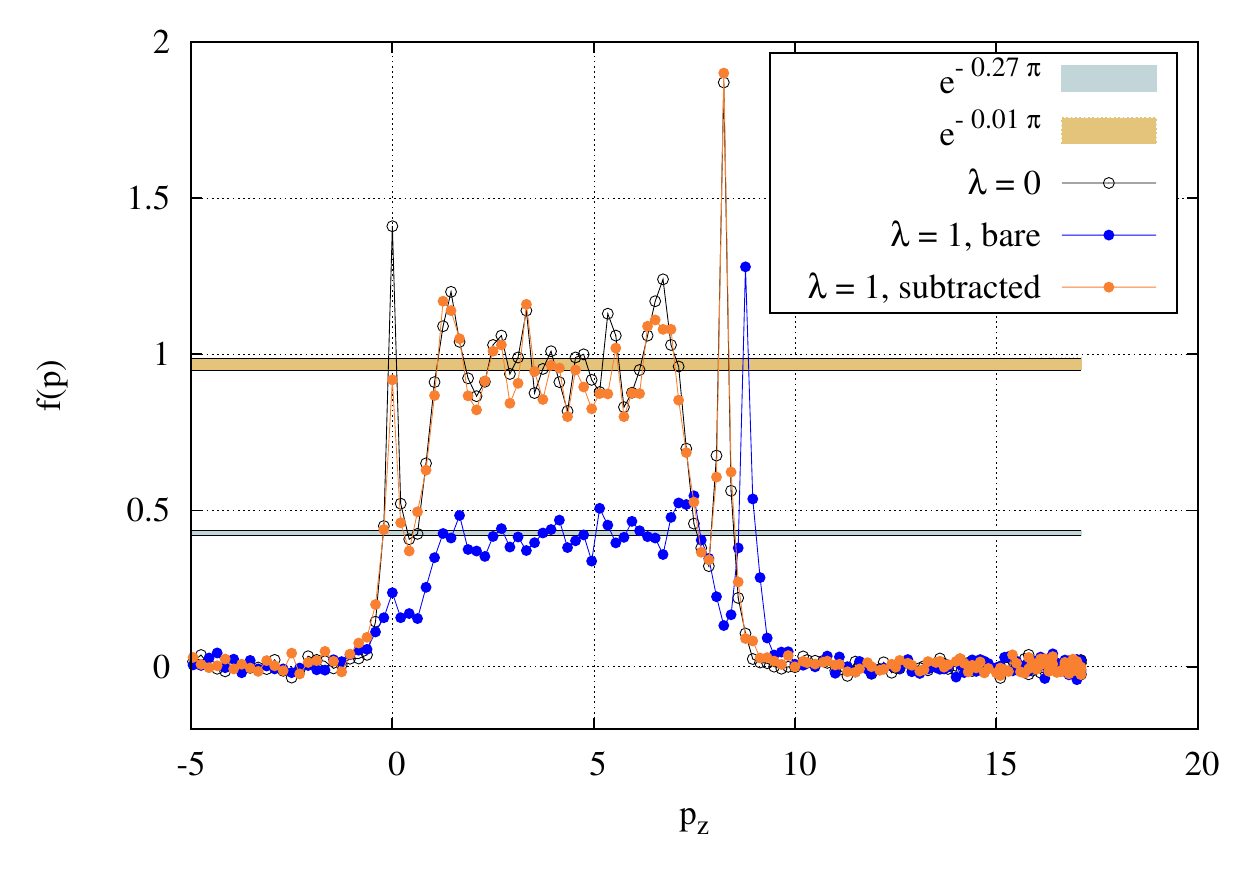}}
  \end{center}
  \caption{\label{fig:ren} $p_z$ spectrum of the produced scalar
    particles at $t=10$, in the non self-interacting case ($\lambda=0$)
    and the self-interacting case ($\lambda=1$), without and with mass
    renormalization. The two horizontal bands indicate the values of
    $\exp(-\pi m^2/eE)$ for $m^2=0.01$ and $m^2=0.27$ (see the text for
    an explanation of these values).}
  \end{figure}
\begin{figure}[htbp]
  \begin{center}
  \resizebox*{6cm}{!}{\includegraphics{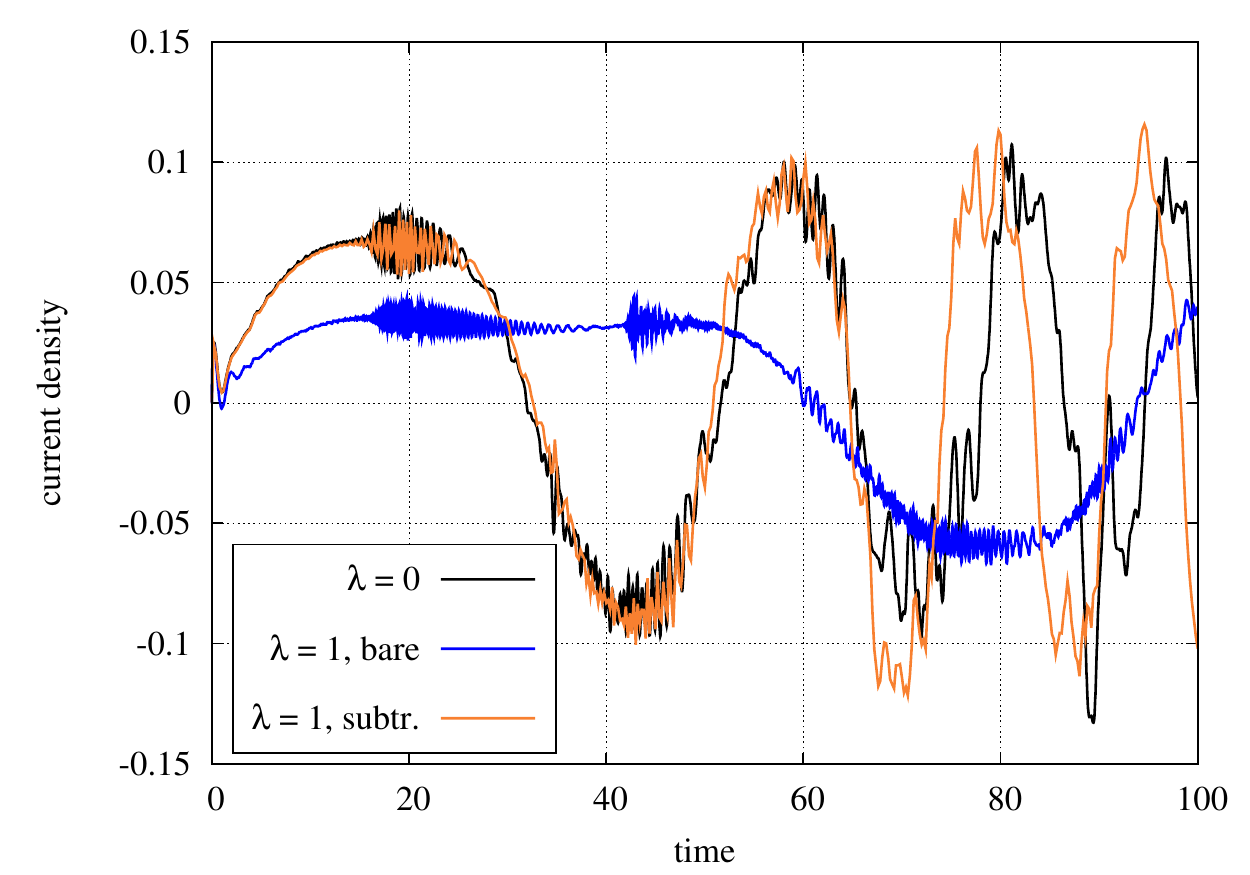}}
  \resizebox*{6cm}{!}{\includegraphics{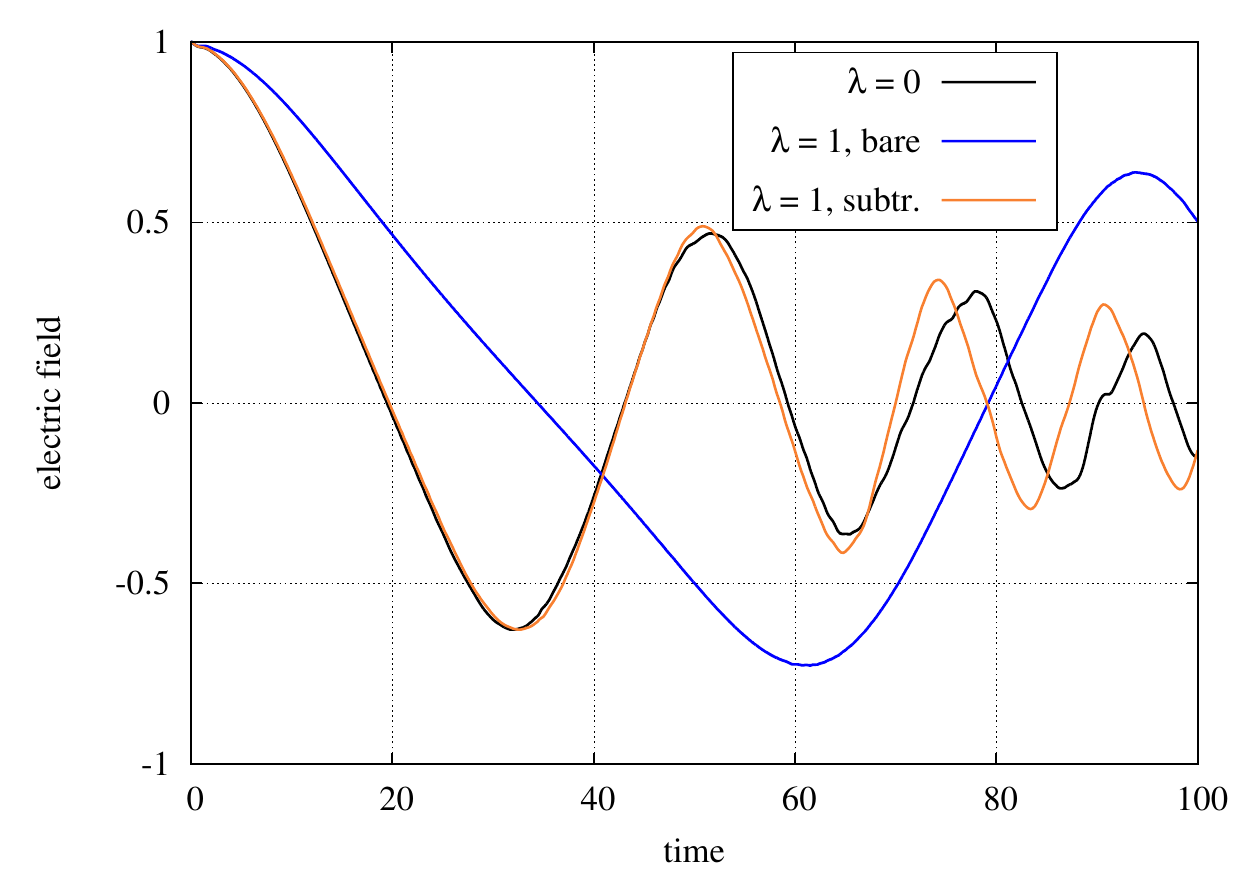}}
  \end{center}
  \caption{\label{fig:cur-l} Time evolution of the electrical current
    density (left) and of the electrical field (right). Curves are shown
    for the non self-interacting case ($\lambda=0$), and for the
    self-interacting case ($\lambda=1$), without and with mass
    renormalization.}
  \end{figure}

  This problem is illustrated in the figure \ref{fig:ren}. The black
  curve, which is almost overlapping with the orange curve,	 
  shows the spectrum of produced particles (at $t=10$,
  i.e. shortly after the external field has been switched on) without
  self-interactions (i.e. with a coupling constant $\lambda=0$). The
  (bare) mass in the Lagrangian, and hence in the classical equation of
  motion, is $m=0.1$.  This curve should be compared to the blue curve,
  where the same computation has been performed with a nonzero
  self-coupling $\lambda=1$, and the same value of the bare mass. We see
  that the particle yield has been considerably reduced. As we shall
  argue, this is a consequence of the fact that these two computations
  correspond to two different values of the renormalized mass.  This is
  an unphysical effect that should be fixed. Indeed, one expects that
  the self-interactions among the scalar fields alter their long time
  evolution (and in particular play a crucial role in their
  thermalization), but should have little physical effects at short
  times. This issue is also visible in the two plots of the figure
  \ref{fig:cur-l}, where the computation at $\lambda=0$ and the bare
  computation at $\lambda=1$ lead to very different results, even at
  short times.

  As explained in the appendix \ref{app:tadpole}, one can remove the
  quadratic divergence coming from the tadpoles by adding a mass
  counterterm $\delta m^2$ in the equation of motion for the classical
  scalar field. This mass counterterm is space-time independent, and can
  thus be computed once for all at the initial time. We define it by
  \begin{equation}
  \delta m^2 \equiv - \lambda\left<\varphi^*(0,\x)\varphi(0,\x)\right>\; ,
  \end{equation}
  which is directly given by eq.~(\ref{eq:init-corr}) and is obviously
  independent of the external electrical field. With the value of the
  self-coupling $\lambda=1$ that we are using, we have
  $\lambda\left<\varphi^*(0,\x)\varphi(0,\x)\right>= 0.26$, which
  is why the calculation done without any mass renormalization gives a
  yield that is well reproduced by $\exp(-\pi m^2/ eE)$ with
  $m^2=0.26+0.01=0.27$ (see the figure \ref{fig:ren}).
  In the figure \ref{fig:ren}, we also show the $p_z$ spectrum obtained
  when this counterterm is added to the classical equation of
  motion. Now, we see that the particle yield is back at the level
  expected for a renormalized mass\footnote{The combination $m^2+\delta
    m^2$ that appears in the equation of motion should now be viewed as
    the bare mass. This bare mass combines with the tadpole that result
    from the Gaussian average, in such a way that the renormalized mass
    is back at the expected value.} $m_{_R}^2=0.01$. Similarly, the
  figure \ref{fig:cur-l} shows that this mass renormalization cures the
  unphysical effect of the self-interactions at short times.

  \section{Conclusions}
  \label{sec:conc}
  Motivated by recent works on thermalization in heavy ion collisions
  using classical statistical field theory, in which one computes 1-loop
  quantum corrections by performing a Gaussian average over the initial
  condition of a purely classical field, we have applied the same method
  to the calculation of the Schwinger mechanism of particle production
  in an external electrical field.

  We have first shown that, at leading order (i.e. at 1-loop), the
  spectrum of particles produced by the Schwinger mechanism can be
  expressed as a path integral over classical fields that have a
  Gaussian ensemble of initial conditions. The 2-point correlation
  function that characterizes this Gaussian distribution is uniquely
  determined by the propagator of small fluctuations on top of the
  external field. This representation of the spectrum is exact at
  1-loop, and is a mere rewriting of the original quantum field theory
  result. Moreover, this formulation of the spectrum leads to a very
  efficient method for the numerical evaluation of the spectrum on a
  lattice.

  Then, by promoting the gauge potential to a dynamical variable, this
  formulation makes easy the inclusion of the back-reaction effects that
  are important for the physical consistency of the model. Indeed, the
  charged particles that are produced via the Schwinger mechanism tend
  to screen the applied external field, thereby reducing progressively
  their production rate. This back-reaction is essential for the
  conservation of total energy (particles + electromagnetic field).

  In the last section, we have studied the possibility of taking into
  account the self-interactions of the produce charged particles, which
  is an essential ingredient for their eventual thermalization. In the
  classical statistical approach, they can be simply included by keeping
  the non-linear interaction term in the classical equation of motion
  for the field. However, since the Schwinger mechanism is very
  sensitive to the value of the mass of the particles, it is important
  to take proper steps in order to renormalize the mass: the naive
  (i.e. without any mass renormalization) inclusion of the
  self-interactions leads to an unphysical --lattice spacing dependent--
  reduction of the charged particle yield, because the self-interactions
  produce large corrections to the mass. The main correction to the mass
  is a quadratic ultraviolet divergence that comes from tadpole
  corrections. We have shown that it can be systematically subtracted in
  the classical statistical framework by adding a mass counterterm in
  the classical field equation of motion.

  To close this paper, we would like to digress with a remark regarding
  the applicability of the classical statistical approach to the
  calculation of observables. The example considered in this paper, as
  well as other quantities previously considered in the literature, is
  an inclusive observable. This means that it measures the expectation
  value of a certain operator (here the particle number operator) in the
  final state of the system, without putting any constraints on this
  final state. This is the reason why these observables can be expressed
  in terms of fields that obey retarded boundary conditions, which in
  turn are amenable to a computation in terms of a Gaussian average over
  their initial conditions. Very little is known about exclusive
  observables, whose definition veto certain final states. The
  archetypal example of this kind of observable would be the
  probability of producing a specific number of charged particles, which
  obviously excludes any final state that does not contain the expected
  number of particles. In some examples, it has been shown that
  exclusive observables can be expressed at leading order in terms of
  classical fields that obey non-retarded boundary conditions
  (e.g. fields that are constrained both at $t=-\infty$ and at
  $t=+\infty$). At least on the surface, it seems very implausible that
  such an observable can be obtained by a classical statistical
  simulation where one performs an average over the {\sl initial}
  conditions of the classical field.

  \section*{Acknowledgements}
  We would like to thank J. Berges, J.-P. Blaizot, T. Epelbaum and B.~Wu
  for useful discussions related to this work. This research is
  supported by the European Research Council under the Advanced
  Investigator Grant ERC-AD-267258. F.G. is supported by Agence
  Nationale de la Recherche project \#~11-BS04-015-01.  N.T. is
  supported by the Japan Society for the Promotion of Science for Young
  Scientists. The numerical part of this work was performed using the
  HPC resources from GENCI-CCRT (Grant t2013056929).

  \appendix

  \section{Tadpoles in classical statistical field theory}
  \label{app:tadpole}
  In this appendix, we show how to deal with the quadratic divergences
  that arise from the tadpoles when we keep the self-interactions in a
  classical statistical simulation. In this appendix, we disregard the
  coupling of the scalars to the gauge fields, since our purpose is to
  discuss an issue related to the scalar self-coupling.  Moreover, in
  order to keep the notations simple, we use continuum notations in this
  appendix. In practical applications to an actual lattice computation,
  all the integrals would become discrete sums.

  Let us illustrate the issue in the case of the simple calculation of
  the expectation value of the field operator itself,
  $\big<\phi(x)\big>$. Before we go further, it is useful to recall the
  Green's formula that relates the classical field $\varphi(x)$ to its
  initial value at $x^0=0$:
  \begin{eqnarray}
  \varphi(x)
  =
  -\frac{\lambda}{2}\int_{y^0>0} \!\!\!
  d^4y\;G_{_R}^0(x,y)\,\varphi^*(y)\varphi^2(y)
  +\int_{y^0=0}\!\!\!d^3\y\; G_{_R}^0(x,y)\stackrel{\leftrightarrow}{\partial^0_y}
  \varphi(0,\y)\; ,
  \label{eq:green}
  \end{eqnarray}
  where $G_{_R}^0(x,y)$ is the free {\sl retarded} Green's function of
  the Dalembertian operator. By iterating the interaction term, one sees
  that the functional dependence of $\varphi$ with respect to its initial
  condition can be represented as a sum of tree diagrams, of the form:
  \setbox1\hbox to 6cm{\hfil\resizebox*{6cm}{!}{\includegraphics{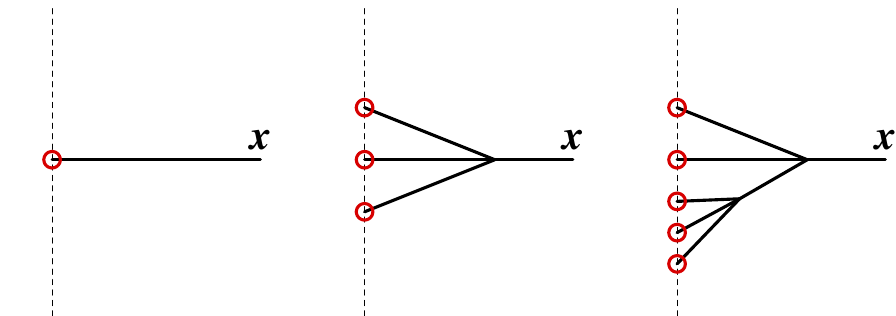}}}
  \begin{equation}
  \varphi(x)=\raise -9.5mm\box1+\cdots
  \label{eq:class_init_tree}
  \end{equation}
  In this equation, we have represented the terms that arise up to the
  second order in the coupling. The vertical dotted line symbolizes the
  initial time surface $y^0=0$, and the red circles the initial value of
  the field $\varphi$ at this initial time. The average over the initial
  field is a Gaussian average; it can be done diagrammatically by
  introducing the following objects
  \setbox1\hbox to 4cm{\hfil\resizebox*{4cm}{!}{\includegraphics{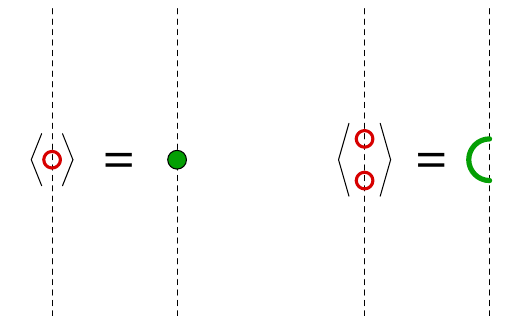}}}
  \begin{equation}
  \raise -10mm\box1
  \end{equation}
  where the green dot represents the central value of the Gaussian
  ensemble\footnote{Assuming a non-zero central value is a slight
    generalization of eq.~(\ref{eq:init-corr}), that allows to have a
    non-zero $\big<\phi(x)\big>$ for the purposes of the discussion in
    this appendix.}, and the green link represents its 2-point
  correlation function. When we apply these diagrammatic rules to the
  expectation value of the field operator, we obtain the following
  contributions \setbox1\hbox to
  10cm{\hfil\resizebox*{10cm}{!}{\includegraphics{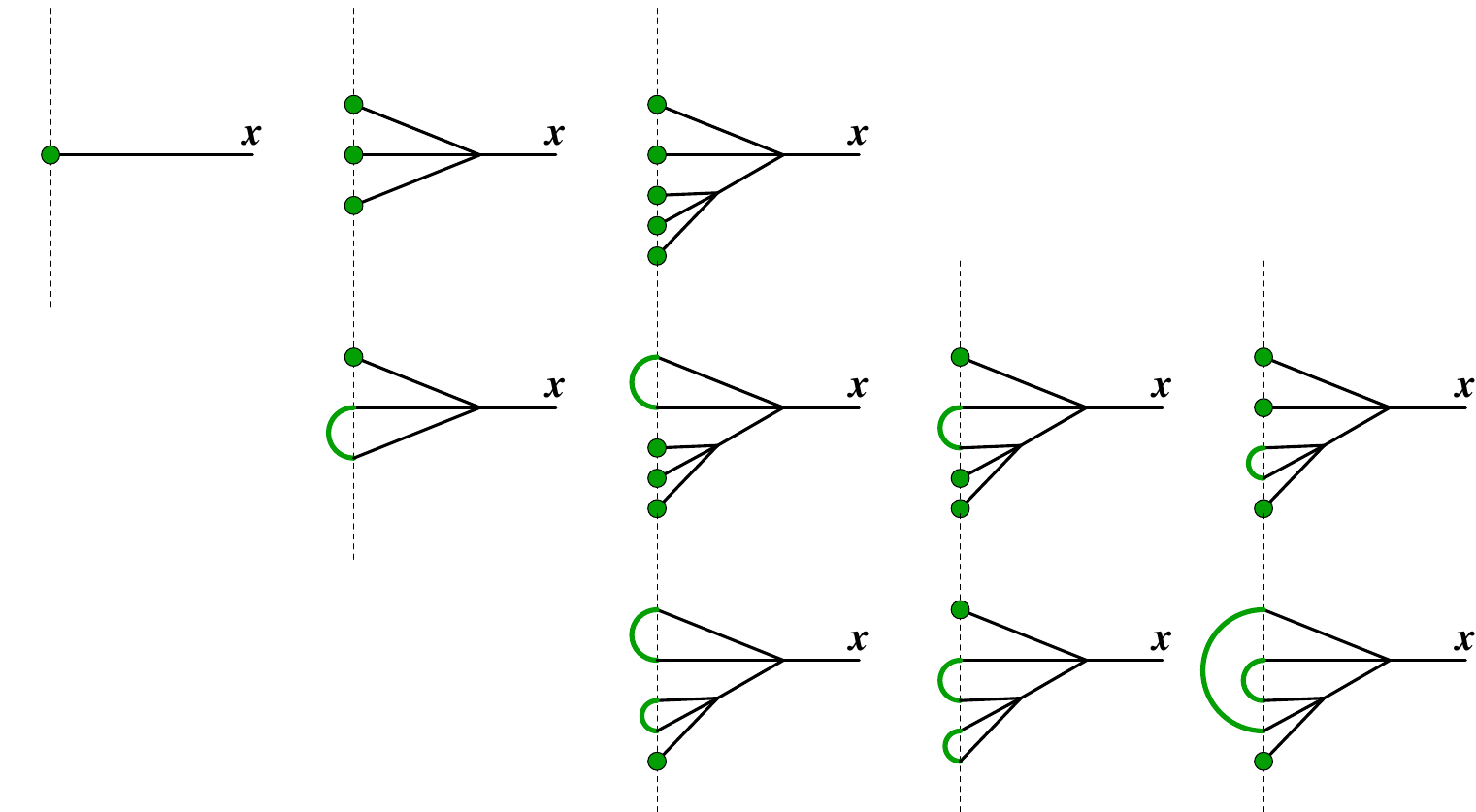}}}
  \begin{equation}
  \left<\phi(x)\right>=\raise -43.5mm\box1
  \label{eq:avg}
  \end{equation}
  In this diagrammatic representation, the green lines and dots
  represent the average over the initial condition, while the black
  lines are genuine (retarded) propagators coming from the subsequent
  time evolution. The first line is the sum of tree level contributions,
  the second line is the 1-loop contribution, etc.. The tree level
  contribution (first line) is nothing but the classical field whose
  initial condition at $x^0=0$ is the central value of the Gaussian
  ensemble, $\varphi_0$. We see that in the classical statistical
  approach, this classical solution is corrected by an infinite set of
  loop corrections.

  The tadpoles are the loops that have the strongest dependence on the
  ultraviolet cutoff. Before going further, it is worth clarifying an
  important point: it is not obvious a priori that the tadpole-like
  subgraphs that appear in the diagrams of eq.~(\ref{eq:avg}) are
  identical to the usual tadpoles encountered in Feynman's perturbation
  theory. Let us demonstrate that they are in fact equal. By using the
  Green's formula (\ref{eq:green}), the tadpoles of eq.~(\ref{eq:avg})
  can be expressed as \setbox1\hbox to
  1.4cm{\hfil\resizebox*{1.4cm}{!}{\includegraphics{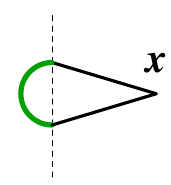}}}
  \begin{eqnarray}
  \raise -6.5mm\box1 \!\!&=&\!\!
  \int\limits_{y^0,z^0=0}\!\! d^3\y d^3\z\;
  \Big[G_{_R}^0(x,y)\stackrel{\leftrightarrow}{\partial^0}_y\Big]
  \Big[G_{_R}^0(x,z)\stackrel{\leftrightarrow}{\partial^0}_z\Big]
  \left<\varphi(0,\y)\varphi(0,\z)\right>
  \nonumber\\
  \!\!&=&\!\!\int\limits_{y^0,z^0=0}\!\! d^3\y d^3\z
  \int\frac{d^4p d^4q}{(2\pi)^8}\frac{1}{(p^2+ip^0\epsilon)(q^2+iq^0\epsilon)}
  \nonumber\\
  &&\qquad\qquad\times
  \Big[e^{-ip\cdot(x-y)}\stackrel{\leftrightarrow}{\partial^0}_y\Big]
  \Big[e^{-iq\cdot(x-z)}\stackrel{\leftrightarrow}{\partial^0}_z\Big]
  \left<\varphi(0,\y)\varphi(0,\z)\right>
  \nonumber\\
  &=&\int\frac{d^3\k}{(2\pi)^3 2k}\; .
  \label{eq:tadpole}
  \end{eqnarray}
  In the second line, we have replaced the retarded propagators by their
  Fourier representation. Then, a straightforward calculation, using
  eqs.~(\ref{eq:init-corr}), leads to the final expression -- which is
  indeed the usual vacuum tadpole.  Given the combinatorics for
  connecting a $\varphi$ and a $\varphi^*$ in the product
  $\varphi^*\varphi^2$, each tadpole will arise with a prefactor
  $\lambda$ in the expansion of eq.~(\ref{eq:avg}), which is also the
  right coupling and symmetry factor.

  It is possible to subtract all the tadpoles that arise during the time
  evolution by adding a mass counterterm $\delta m^2$ in the classical
  equation of motion, that becomes
  \begin{equation}
    \left(\square+m^2+\delta m^2\right)\varphi+\frac{\lambda}{2}\varphi^*\varphi^2=0\;.
  \label{eq:eom-renorm}
  \end{equation}
  The effect of this counterterm in the equation of motion is to insert
  in its solution a mass counterterm in every place where a tadpole is
  allowed to appear. It therefore adds the following contributions to
  eq.~(\ref{eq:avg}) \setbox1\hbox to
  6cm{\hfil\resizebox*{6cm}{!}{\includegraphics{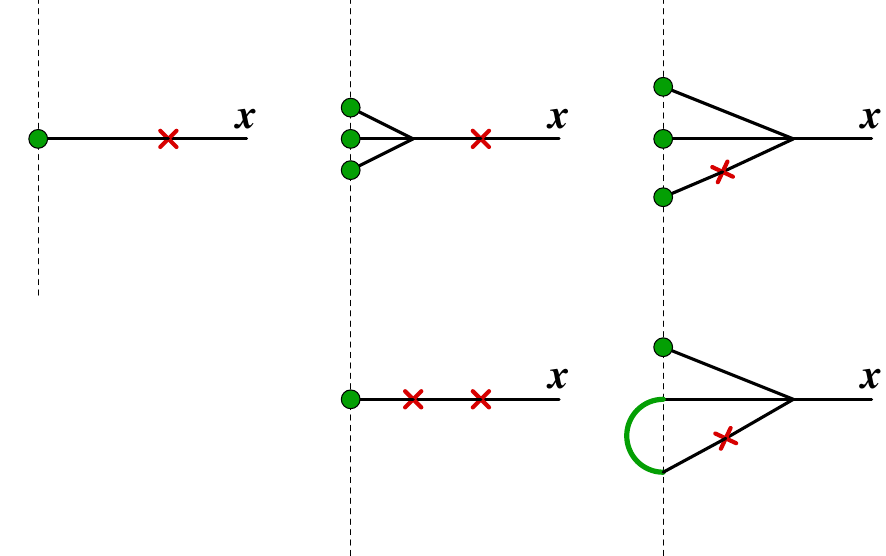}}}
  \begin{equation}
  \delta\left<\varphi(x)\right>=\raise -27.5mm\box1
  \end{equation}
  where the red cross denotes the mass counterterm. The outcome is that
  the tadpoles are systematically cancelled by this procedure if one
  tunes the counterterm to precisely cancel the quadratic divergence of
  the tadpole,
  \begin{equation}
  \delta m^2 +\lambda \int\frac{d^3\k}{(2\pi)^3 2k} < \infty\; .
  \label{eq:ren}
  \end{equation}
  Eq.~(\ref{eq:ren}) only specifies the divergent part of the mass
  counterterm; it also has a finite part that should be adjusted in
  order to have the desired renormalized mass. Note also that the
  inclusion of this counterterm in the equation of motion only subtracts
  the quadratic divergences, possibly leaving a residual logarithmic
  dependence on the lattice spacing.


\end{document}